\documentclass[aps,pre,10pt,notitlepage,nofootinbib,superscriptaddress]{revtex4-2}

\usepackage{amsmath,amssymb,graphicx,xcolor,tabularray,accents,hyperref}

\newcommand{\vecX}{{\bf X}}
\newcommand{\vecj}{{\bf j}}
\newcommand{\vecn}{{\bf n}}
\newcommand{\veco}{{\mathbf{0}}} 
\newcommand{\vecq}{{\bf q}}
\newcommand{\vecr}{{\bf r}}
\newcommand{\vecv}{{\bf v}}
\newcommand{\vecx}{{\bf x}}
\newcommand{\z}{Z}
\newcommand{\prob}{\Pr}
\newcommand{\dx}{\overline{d\vecx}}
\newcommand{\dX}{\overline{d\vecX}}
\newcommand{\vecG}{\underaccent{\bar}{G}}
\newcommand{\bbo}{\underaccent{\bar}{0}} 
\newcommand{\vecb}{\underaccent{\bar}{b}}
\newcommand{\vecf}{\underaccent{\bar}{f}}
\newcommand{\vecg}{\underaccent{\bar}{g}}
\newcommand{\vecJ}{\underaccent{\bar}{\phi}}
\newcommand{\bmu}{\underaccent{\bar}{\mu}}
\newcommand{\bpsi}{\underaccent{\bar}{\psi}}
\newcommand{\vecA}{{\cal A}}
\newcommand{\vecB}{{\cal B}}
\newcommand{\bphi}{{\cal C}} 
\newcommand{\vecD}{{\cal D}}
\newcommand{\vecM}{{\cal M}}
\newcommand{\calo}{{\cal O}} 
\newcommand{\blam}{{\cal U}} 
\newcommand{\brho}{\bar{\rho}}
\newcommand{\Nmean}{\langle N\rangle}

\newcommand{\baa}{\overline{a^2}}
\newcommand{\mf}{M} 
\newcommand{\mq}{\Omega} 
\newcommand{\mt}{\Xi} 
\newcommand{\eg}{{e.g., }}
\newcommand{\ie}{{i.e., }}
\newcommand{\tr}{{\rm tr}}

\begin{document}

\title{Resolving entropy contributions in nonequilibrium transitions}



\author{Benjamin Sorkin}

\affiliation{School of Chemistry and Center for Physics and Chemistry of Living Systems, Tel Aviv
  University, 69978 Tel Aviv, Israel}

\author{Joshua Ricouvier}

\affiliation{Department of Chemical and Biological Physics, Weizmann Institute of Science, 76100 Rehovot, Israel}

\author{Haim Diamant}

\affiliation{School of Chemistry and Center for Physics and Chemistry of Living Systems, Tel Aviv
  University, 69978 Tel Aviv, Israel}

\author{Gil Ariel}
\email{arielg@math.biu.ac.il}

\affiliation{Department of Mathematics, Bar-Ilan University, 52000
  Ramat Gan, Israel}

\begin{abstract}

We derive a functional for the entropy contributed by any microscopic degrees of freedom as arising from their measurable pair correlations. Applicable both in and out of equilibrium, this functional yields the maximum entropy which a system can have given a certain correlation function. When applied to different correlations, the method allows us to identify the degrees of freedom governing a certain physical regime, thus capturing and characterizing dynamic transitions. The formalism applies also to systems whose translational invariance is broken by external forces and whose number of particles may vary. We apply it to experimental results for jammed bidisperse emulsions, capturing the crossover of this nonequilibrium system from crystalline to disordered hyperuniform structures as a function of mixture composition. We discover that the cross-correlations between the positions and sizes of droplets in the emulsion play the central role in the formation of the disordered hyperuniform states. We discuss implications of the approach for entropy estimation out of equilibrium and for characterizing transitions in disordered systems.

\end{abstract}

\maketitle

\section{Introduction}
\label{sec_intro}

Macroscopic states of matter differ by the statistics of their microscopic degrees of freedom (DOFs). Moreover, the physical mechanism underlying a change of state is reflected in the roles played by the different DOFs. One way to quantify the contributions of particular DOFs (or the interactions between them) to the change of state, is through their effect on the entropy. Unlike temperature and pressure, whose consistent definitions, and thus measurement, are problematic out of equilibrium~\cite{CasasVazquez2003,Cugliandolo2011,Solon2015}, the relation between entropy and information content is believed to hold for any steady state, regardless of whether the system is at thermodynamic equilibrium or not.

According to Shannon's definition~\cite{Shannon1948}, the information entropy is given by
\begin{subequations}
\label{eq:shan}
\begin{equation}
  H[P_s] = -\sum_s P_s \ln P_s,
\end{equation}
$P_s$ being the probability distribution of finding the system in the discrete microstate $s$. The continuous form, relevant in many physical systems, is
\begin{equation}
  H[P(\vecX)] = -\int d\vecX\, P(\vecX) \ln P(\vecX),\label{eq:cont_shan}
\end{equation}
\end{subequations}
with $P(\vecX)$ being the probability density function of finding the system around the phase-space point $\vecX$.

Applying Eqs.~\eqref{eq:shan} directly and, furthermore, resolving the contributions of the different DOFs, is usually impractical; out of equilibrium, the required knowledge of $P_s$ (and even more so, $P(\vecX)$), is prohibitively hard to acquire by sampling~\cite{Kozachenko1987,Beirlant1997review,Darbellay1999,Paninski2003,Stowell2009,Lord2018,Ariel2020}. Recently, several algorithms aiming to estimate entropy of physical systems have been demonstrated, based on data compression~\cite{AvineryPRL2019,MartinianiPRX2019,Zu2020,feldman2003comp,melchert2015comp} or recursive mutual-information estimation \cite{NirPNAS2020}. Since such methods still rely on sampling of microscopic configurations, they are bound to suffer from undersampling for real materials due to the high dimensionality and continuity of phase space.

Given that the detailed $P(\vecX)$ is practically unattainable, we propose to follow the trusted path of statistical mechanics and consider instead lower-dimensional marginal distributions, namely, correlation functions (CFs)~\cite{Book:Hansen}.
Although it may seem like a severe compromise, this coarse-graining has three important advantages. First, spatial CFs are robustly measurable in experiments and simulations, without extensive sampling of microscopic states. Second, they provide a useful characterization of the structure of matter, both in and out of equilibrium~\cite{Book:Hansen,cavagna2016CF,ro2021disorder,pearce2021orientational,sanchez2012spontaneous}. Third, by supplying different CFs, we will be able to discern their contributions to the entropy and determine the dominant factor underlying the changes in the system.

An illustrative example is the isotropic-to-nematic transition in a system of hard rods~\cite{OnsagerNematics,deGennesBook}. In this transition the physical mechanism is known, being the competition between the translational and rotational DOFs over the available phase-space volume. Resolving the different entropy contributions from the CF of each DOF, and possibly also their cross-correlation, would clarify this mechanism were it unknown, and indicate the transition point at which the rotational DOFs lose in the entropy competition and become ordered. The contribution of the orientations CF would change considerably between the two phases, while that of the density CF (\ie the structure factor) would hardly change. Later we demonstrate this idea in a system whose underlying physics is much less understood, and whose transitions are less evident in the CFs {\it per se}.

The goal of the present work is to establish a theoretical framework for the above. We wish to estimate the entropy associated with a two-point CF of arbitrary nature. Our strategy is as follows. (a) We look for the information content of a given CF, i.e., the ``entropy cost" of knowing that CF. We do so by finding the distribution $P(\vecX)$ that maximizes the Shannon entropy, Eq.~(\ref{eq:shan}), under the constraint that it should yield the given two-point CF. Substituting it back into Eq.~(\ref{eq:shan}) gives the information content, \ie the entropy bound. Obtaining this expression involves a Gaussian approximation (see Appendix~\ref{appendix:gencor}). A proof of concept for this step in the simplest case, where the DOFs are the positions of point-like particles and the corresponding two-point CF is the structure factor, was presented in Ref.~\cite{ArielPRE2020}. Here we extend the formalism to the general case of any DOFs and their CFs.  (b) Imposing a set of measured CFs, we obtain a series of entropy upper bounds. Comparison of the different bounds points at the DOFs which dominate the physical behavior under consideration, and their interaction. The DOFs are completely arbitrary, and may include labels of particle species, their orientation, velocities, sizes, etc. (see Table~\ref{table:corr}). (c) A tighter upper bound can be obtained by imposing several CF constraints at once. This allows us to isolate the contributions of cross-correlations to the underlying physics. 

The formalism can be generalized further to include additional constraints, such as imposed inhomogeneity and fluctuations in the number of particles (see Appendix~\ref{appendix:inhom}). In addition, as a byproduct, we provide an alternative route for entropy estimation. By integrating out the unavailable statistics we bypass the undersampling problem. 

Indeed maximum-entropy methods have proved useful for obtaining effective interaction models for complex, out-of-equilibrium systems, in both experiments (\eg~\cite{bialek2012,shemesh2013high,asor2019assembly}) and simulations (\eg~\cite{bialek2012,info2003network,cavagna2014ent}). To date, such methods are applicable only for  maximization over a small set of effective parameters~\cite{bialek2012,shemesh2013high,cavagna2014ent,info2003network} and in the particular case of the structure factor~\cite{Chakrabarty2011,ArielPRE2020,Zhang2020}.

The paper is constructed as follows. Section~\ref{sec_gen} gives the main result\,---\,the entropy functional of a given CF\,---\,and describes how it is implemented. In Sec.~\ref{sec_experiment} we apply the formalism to  experimental results obtained for a confined emulsion of polydisperse droplets~\cite{ricouvier2017optimizing}. In Sec.~\ref{sec_discussion} we summarize our findings and discuss the many potential applications of the method. We defer more technical information to the Appendixes and Supplemental Material (SM)~\cite{SM}. In Appendix~\ref{appendix:inhom} we present the required adjustments to the central result when translational symmetry is broken, \eg by an external field. The detailed derivation of the main result is given in Appendix~\ref{appendix:gencor}, for the case where the CF is the only constraint. In the SM we derive the adjusted entropy bound in the presence of imposed inhomogeneity, and the treatment of a variable number of particles (grand-canonical ensemble). We also provide the ``recipe'' for how the CFs in the experiments were constructed, along with more experimental snapshots and figures of the CFs. Finally, the SM presents a toy model demonstrating the failure of various computational methods for entropy estimation in the case of a variable number of particles.

\section{General formalism}
\label{sec_gen}

The system of interest contains the locations of $N$ particles (their centers of mass (CMs)) in $d$ dimensions, $\{\vecr_{n=1\ldots N} \in \mathbb{R}^{d}\}$, and possibly an additional set of DOFs that we denote by $\{j_{k,n=1\ldots N}\}$. The index $k$ marks the scalar components of all additional DOFs, and we abbreviate $\vecj_n=(j_{1,n},j_{2,n},\ldots)$. For example, in the case of rod-like particles, $\vecj_n$ may include the components of the solid angle, CM velocity, and angular velocity of the $n$th rod. The indices may also mark the type of particle in a many-component mixture. Another important example are particles, such as polyatomic molecules, which are made of subparticles (atoms) arranged in certain spatial conformations. See Table~\ref{table:corr} for examples of correlations between such DOFs. 

To keep the analysis as general as possible, we define the field
\begin{subequations}
\label{eq:posfields}
\begin{equation}
    \vecJ(\vecr)=\sum_{n=1}^N\delta(\vecr-\vecr_n)\vecf(\vecj_{n}), \label{eq:realf}
\end{equation}
where $\vecf$ is any function (for simplicity, in the form of a column vector) of the additional DOFs $\vecj_n$. For example, for the particle-density field $\vecf=1$, for the orientation vector field $\vecf=\hat{\vecn}$, and for the nematic tensor field~\cite{deGennesBook}, $\mathbf{Q}=(d\hat{\vecn}\hat{\vecn}^T-\mathbf{I})/(d-1)$, $\vecf$ consists of the independent components of $\mathbf{Q}$. We denote the total number of components within $\vecf$ and $\vecJ(\vecr)$ by $\mf$ (\ie $\vecf,\vecJ(\vecr)\in\mathbb{C}^\mf$).

We now construct the two-point CF associated with the field of Eq.~\eqref{eq:realf}. First, for convenience, the field is Fourier-transformed into
\begin{equation}
    \vecJ(\vecq)=\sum_{n=1}^Ne^{-i\vecq\cdot\vecr_n}\vecf(\vecj_{n}), \label{eq:q-f}
\end{equation}
\end{subequations}
where the uniform ($\vecq=\veco$) mode is excluded~\footnote{Equations~\eqref{eq:posfields} define position-dependent CFs, but our formalism is not restricted to spatial coordinates. For example, the DOFs may be only the orientations, and the Fourier transform would be replaced by a decomposition into spherical harmonics.}. The CF is a $\mf\times\mf$ matrix, defined as
\begin{subequations}
\label{eq:corrnorm}
\begin{eqnarray}
    \bphi(\vecq)=\vecA^{-1}\langle N^{-1}\vecJ(\vecq)\vecJ^\dagger(\vecq)\rangle,
    \label{eq:homcorr}
\end{eqnarray}
where $\langle\cdot\rangle$ denotes an ensemble average. Note that we include $N$ within the brackets in case the number of particles is not fixed. The normalization matrix $\vecA$ ensures that, for uncorrelated fields (ideal gas), $\bphi(\vecq)=\mathcal{I}$ for all $\vecq$, where $\mathcal{I}$ is the $\mf\times \mf$ unit matrix. Explicitly, $\vecA$ is the second-moment matrix of $\vecf(\vecj)$, given by
\begin{equation}
    \vecA=\langle\vecf(\vecj)\vecf^\dagger(\vecj)\rangle=\int d\vecj  \prob(\vecj)\vecf(\vecj)\vecf^\dagger(\vecj),\label{eq:norm}
\end{equation}
\end{subequations}
where $\prob(\vecj)$ is the single-particle marginal distribution of the DOFs $\vecj$. Note that the full distribution is not required (only the second moment of $\vecf$ is), and that $\vecj$ may have either annealed or quenched statistics.
The definitions above hold in more general settings in which $\vecf$ depends explicitly on $\vecq$. In fact, the only condition on $\vecf$ is that $\vecA$ of Eq.~\eqref{eq:norm} is invertible.

Since the CFs are measurable, \eg by scattering or simulations, the $\vecq$ modes are naturally discretized by the system size. While their number is infinite in principle, only a finite number of $\vecq$ modes, $\mq=\sum_{\vecq\ne\veco}1$, are measured in practice. Constraining such a subset of modes will merely loosen the entropy bound. The limit of continuous $\vecq$ will be addressed below as well. 

Let us clarify again the notation used above. The bold symbols are physical vectors, tensors, etc. (\eg $\vecr$, $\mathbf{Q}$). The underlined objects ($\vecf$ and $\vecJ(\vecq)$) are column vectors in $\mathbb{C}^\mf$. Lastly, the calligraphic objects ($\bphi(\vecq)$, $\vecA$, and $\mathcal{I}$) are $\mf\times \mf$ matrices. See again Table~\ref{table:corr} for a few specific examples.

\begin{table}[ht]
\begin{tblr}{|c|c|c|c|}
\hline
Field & $\vecj_n$ & $\vecf(\vecj)$ & $\bphi(\vecq)$ \\
\hline
\hline
Density & None & 1 & $\bphi\equiv S(\vecq)= \left\langle\frac{1}{N}\sum_{n,m}e^{-i\vecq\cdot\vecr_{nm}}\right\rangle$ \\
\hline
Velocity & $\vecv_n$ & $\vecv$ & $\bphi_{\alpha\beta}= (\vecA^{-1})_{\alpha\gamma} \left\langle\frac{1}{N}\sum_{n,m}e^{-i\vecq\cdot\vecr_{nm}}v_{\gamma,n}v_{\beta,m}\right\rangle$ \\
\hline
Orientation & $\hat{\vecn}_n$ & $\hat{\vecn}$ & $\bphi_{\alpha\beta}
= \left\langle\frac{d}{N}\sum_{n,m}e^{-i\vecq\cdot\vecr_{nm}}\hat{n}_{\alpha,n}\hat{n}_{\beta,m}\right\rangle$ \\
\cline{1,3,4}
Nematic tensor && $\{\mathbf{Q}_{\alpha\geq\beta}(\hat{\vecn})\}$ & $\bphi_{ij}=(\vecA^{-1})_{ik} \left\langle\frac{1}{N}\sum_{n,m}e^{-i\vecq\cdot\vecr_{nm}}Q_k(\hat{\vecn}_n)Q_j(\hat{\vecn}_m)\right\rangle$ \\
\hline
Mixture composition & $t_n$ & $(\delta_{t1}, \delta_{t2})^T$ & $\bphi_{ij}\equiv \mathcal{S}_{ij}(\vecq)=\left\langle\frac{1}{N_i}\sum_{n=1}^{N_i}\sum_{m=1}^{N_j}e^{-i\vecq\cdot\vecr_{nm}}\right\rangle$ \\
\hline
Volume fraction & $a_n$ & $a$ & $\bphi\equiv \chi(\vecq)=\frac{1}{\baa}\left\langle\frac{1}{N}\sum_{n,m}e^{-i\vecq\cdot\vecr_{nm}}a_na_m\right\rangle$ \\
\hline
\end{tblr}
\caption{Examples of possible fields and their corresponding correlation functions, $\bphi(\vecq)$. In all examples, $\vecr_{nm}=\vecr_n-\vecr_m$ is the interparticle separation. The $n,m$ summations are over all $N$ particles. Greek indices correspond to the $d$ spatial axes. Einstein's summation rule is applied. The normalization matrices $\vecA$ are computed using Eq.~\eqref{eq:norm}. The unit vector $\hat{\vecn}$ is the solid angle. The nematic tensor is $\mathbf{Q}=(d\hat{\vecn}\hat{\vecn}^\dagger-\mathbf{I})/(d-1)$. Since $\mathbf{Q}$ is symmetric, we rearrange its $M=d(d+1)/2$ unique components in a column vector. For binary mixtures, $t_n=1,2$ denotes the type of the $n$th particle, and the structure factor adopts a matrix form, $\mathcal{S}_{ij}(\vecq)$, where $i,j=1,2$ are particle-type labels and $N_i$ is the number of  $i$-type particles. In the volume-fraction field, $a_n$ is the $n$th particle volume, and $\baa=\int da\prob(a)a^2$ is the second moment of the single-particle volume distribution. The correlation function $\chi(\vecq)$ is known as the spectral density.}
\label{table:corr}
\end{table}

Now that the mathematical structure of the constraints has been defined, we give our central result\,---\,an upper bound on the thermodynamic entropy $S$,
\begin{equation}
    \frac{S-S^\mathrm{id}}{\Nmean}\leq h_{\rm ex}.
    \label{eq:bound}
\end{equation}
Here, $\Nmean$ is the mean number of particles, $S^{\mathrm{id}}$ is the entropy of the ideal gas, and $h_\mathrm{ex}$ is the excess entropy (per particle) of the measured two-point CF, $\bphi(\vecq)$ (its information content), which we find to be approximately
\begin{eqnarray}
    h_{\rm ex}[\bphi]&=&\frac{1}{2\langle N\rangle} \sum_{\vecq\ne\veco}\tr\left[\ln\left(\bphi(\vecq)\right)+\mathcal{I}-\bphi(\vecq)\right].\label{eq:dischom}
\end{eqnarray}
The matrix logarithm appearing in Eq.~\eqref{eq:dischom} is handled according to $\tr[\ln(\cdot)]=\ln[\det(\cdot)]$. This expression is also applicable to cases where the number of particles is not fixed. The inequality in Eq.~\eqref{eq:bound} comes from the fact that the CF may not constitute the entire information encoded in the system's state (as in systems with many-body interactions or out of equilibrium). The bound is  approximate due to a leading-order (Gaussian) approximation which has been employed (see Appendix~\ref{appendix:gencor}). 

The transition to continuous $\vecq$ is done through $\sum_{\vecq}(\cdot)=\nu\int d\vecq(\cdot)$, where $\nu=V/(2\pi)^d$ is the density of $\vecq$ modes, and $V=\int d\vecr$ is the system's volume. The continuous version of Eq.~\eqref{eq:dischom} is then
\begin{eqnarray}
    h_{\rm ex}[\bphi]&=&\frac{1}{2(2\pi)^d\brho} \int_{\vecq\ne\veco} d\vecq\,\tr\left[ \ln\left(\bphi(\vecq)\right)+\mathcal{I}-\bphi(\vecq)\right],
    \label{eq:cont_hom_ent}
\end{eqnarray}
where $\brho=\Nmean/V$ is the mean density \footnote{In certain cases, as in the case of  the structure factor, the CF includes a contribution of order $N$ at $\vecq = \veco$, which is why this mode is removed from the summation or integration. They must begin from $|\vecq|\to0^+$.}.

The detailed derivation of Eq.~\eqref{eq:dischom} is given in Appendix~\ref{appendix:gencor}. Although the setup above may seem complicated, it reflects the wide applicability of the approach. In practice, once the commonly used CFs (\eg structure factor, orientations or velocity CFs; see Table~\ref{table:corr}) are found, they can simply be inserted into Eq.~\eqref{eq:dischom} to obtain useful entropy bounds. This will become clearer below when we apply Eq.~\eqref{eq:dischom} to a specific system.

Equation~\eqref{eq:dischom} can be extended to cases where an external field is present. For example, with $\vecf=1$ (density field), the steady-state density of particles may be nonuniform (\eg due to gravity). We can include such a nonuniform steady profile, $\vecG(\vecq)=\langle N^{-1/2}\vecJ(\vecq)\rangle$, as another measurable constraint. In addition, due to the breaking of translational symmetry, the CF defined in Eq.~\eqref{eq:homcorr} would cease to be $\vecq$ diagonal and become $\bphi(\vecq,\vecq')=\vecA^{-1}\langle N^{-1}\vecJ(\vecq)\vecJ^\dagger(\vecq')\rangle$. These modifications are treated in Appendix~\ref{appendix:inhom}. The resulting bound, extending Eq.~\eqref{eq:dischom}, is given in Eq.~\eqref{eq:hent}.

\section{Application to bidisperse mixtures}
\label{sec_experiment}

Dense random packings of particles have been a central issue in statistical physics for years~\cite{TorquatoBook}. Recently, disordered random packings close to the jamming point, referred to as maximally random-jammed, have been found to exhibit hyperuniformity~\cite{ZacharyPRL2011,JiaoPRE2011,ricouvier2017optimizing,WilkenPRL2021}\,---\,strong suppression of density fluctuations over large distances~\cite{torquato:review}. In two dimensions, densely packed monodisperse disks tend to locally arrange into hexagonal crystals. However, defects are usually present in the crystal structure, randomly distributed over space. A small degree of polydispersity limits the size of the crystalline domains and reduces the presence of rattlers. Hence, a densely packed mixture of bidisperse disks is expected to transition, as a function of the number fraction ($\textit{NF}$\,) and size ratio ($\textit{SR}$\,) of the two species, between mostly crystalline structures and  disordered structures.

Experiments on jammed quasi-2D emulsions of bidisperse droplets, confined and flattened between two surfaces in a microfluidic channel, have demonstrated this  optimization~\cite{ricouvier2017optimizing}. They showed, moreover, that the increase in disorder came together with stronger hyperuniformity. In this section we revisit the data from those experiments in light of the present formalism, to obtain the entropy contributions of the droplets' different DOFs (position and size), and examine how these contributions change along the transition.

First, we briefly describe the experimental setup; an extended description of the experimental devices and conditions can be found in Ref.~\cite{ricouvier2017optimizing}. Standard soft photolithography and replica-molding techniques are used to create the microfluidic device~\cite{duffy1998rapid}. The pressures applied at the inlets of two T junctions allow a precise control over the droplet sizes and their production rates. Downstream, a microfluidic mixer randomizes the entrance of the droplets into the collection chamber. The chamber height ($10$~$\mu\mathrm{m}$) forces the droplets to adopt a pancake-like shape, and they  assemble into 2D configurations.
Their deformability and the pressure ensure that the assembly is close to a jammed configuration. Images are taken inside the observation chamber every 30 seconds in order for all the droplets to be renewed. Thus, the snapshots depict uncorrelated configurations. A total of 30 such images are taken for every set of parameters. For each set we verify that (a) the two populations of droplets are well mixed, (b) the droplets maintain a circular shape, and (c) the Fourier transform of each image has a circular symmetry.

Here, we focus on two sets of data. In the first, the size ratio is kept fixed at $\textit{SR}\equiv R_\mathrm{small}/R_\mathrm{big} \simeq 0.8 ~ (0.7-0.82)$, while the number fraction, $\textit{NF}\equiv N_\mathrm{big}/(N_\mathrm{big}+N_\mathrm{small})$, is varied. In the second set, the number fraction is kept fixed at $\textit{NF}\simeq 0.5 ~ (0.4-0.6)$, while the size ratio is varied.
For comparison, we also study by simulation the corresponding 2D bidisperse jammed configurations of hard disks. The simulations apply the freely available code based on the Lubachevsky-Stillinger algorithm~\cite{skoge2006packing,atkinson2016static}. 

\begin{figure}
  \raisebox{7.3em}{$\textit{SR}=0.80$}\raisebox{6em}{\hspace{-4.55em}$\textit{NF}=0.94$}\; \raisebox{2em}{\includegraphics[width=0.2\linewidth]{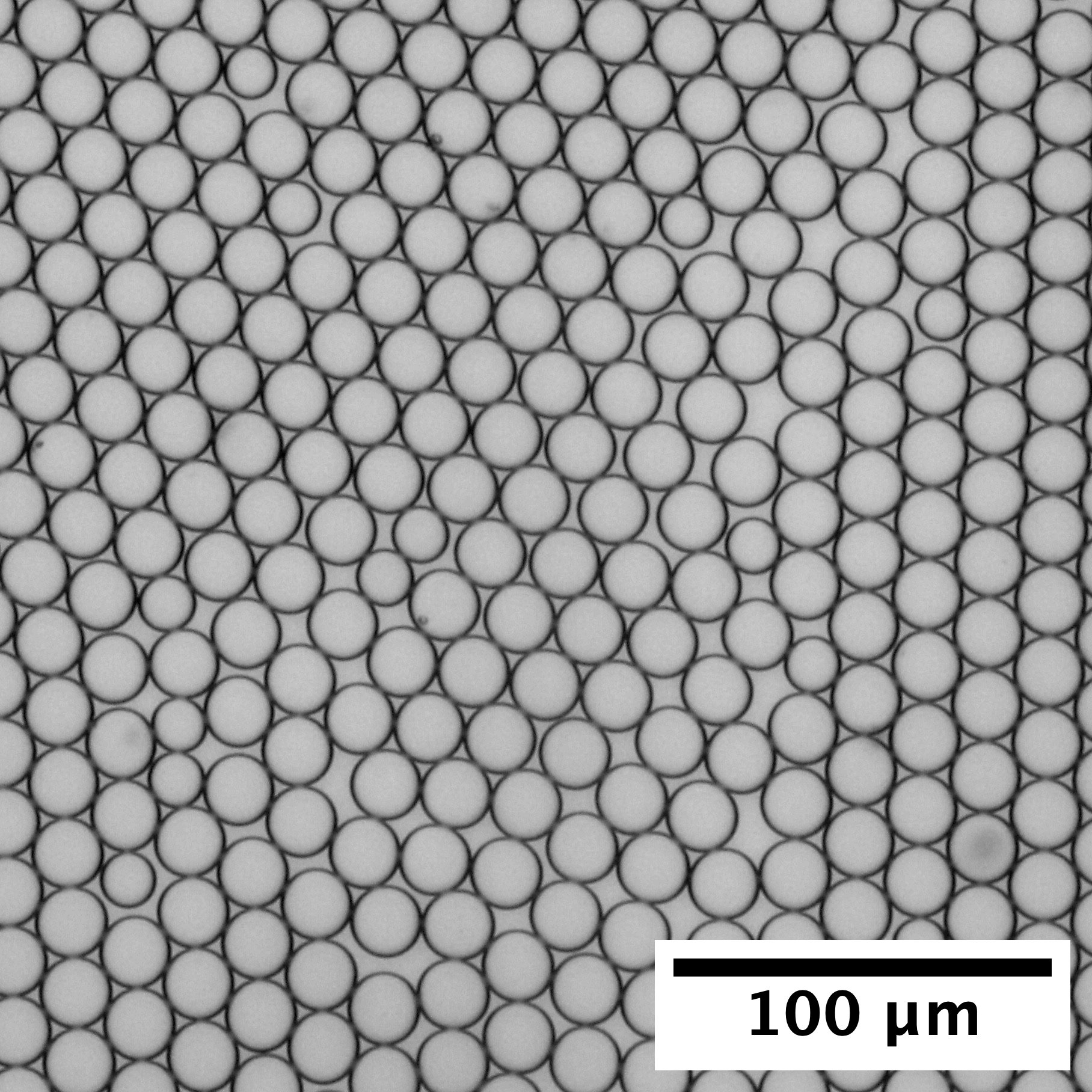}}\includegraphics[width=0.36\linewidth]{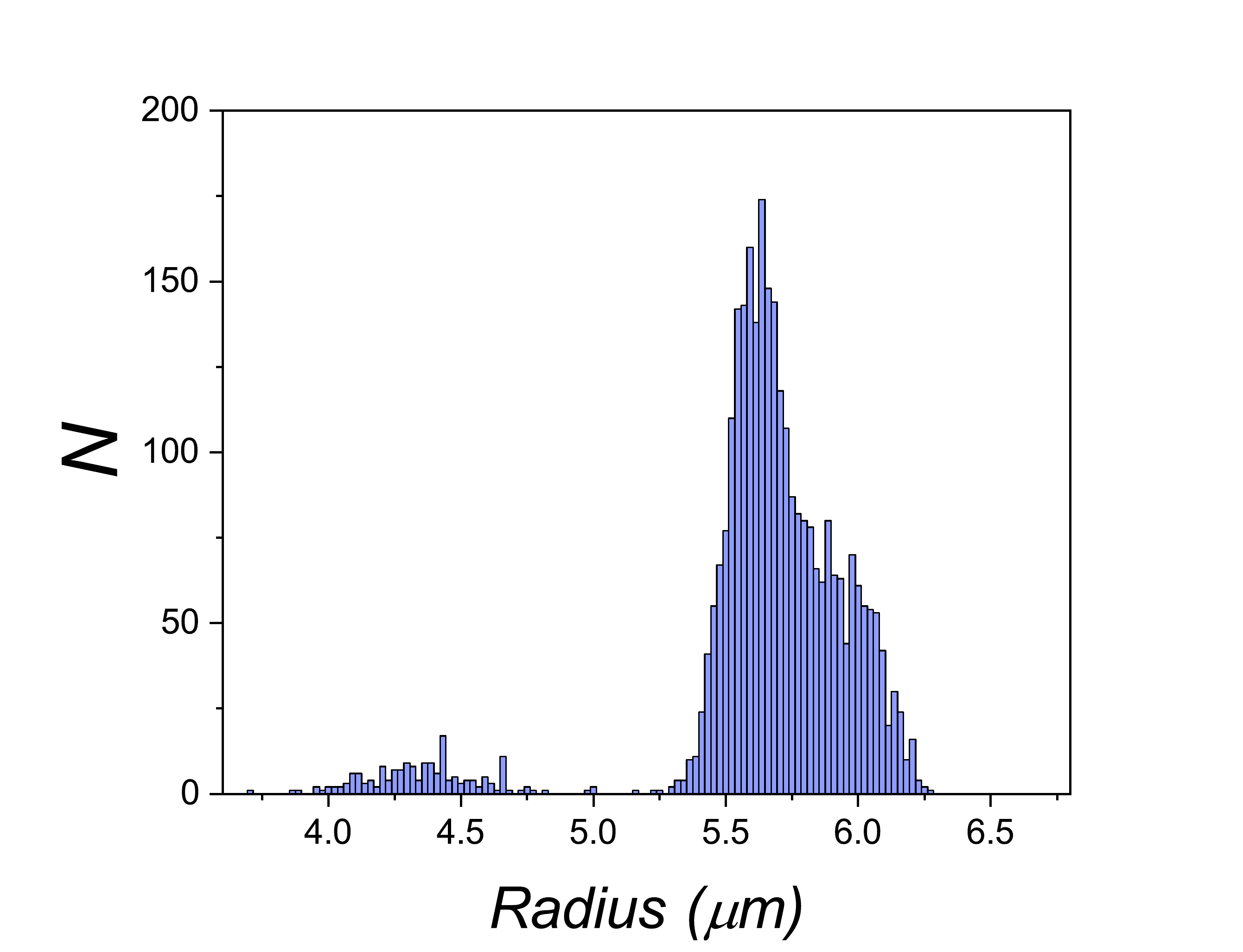}\hspace{-2.6em}\includegraphics[width=0.36\linewidth]{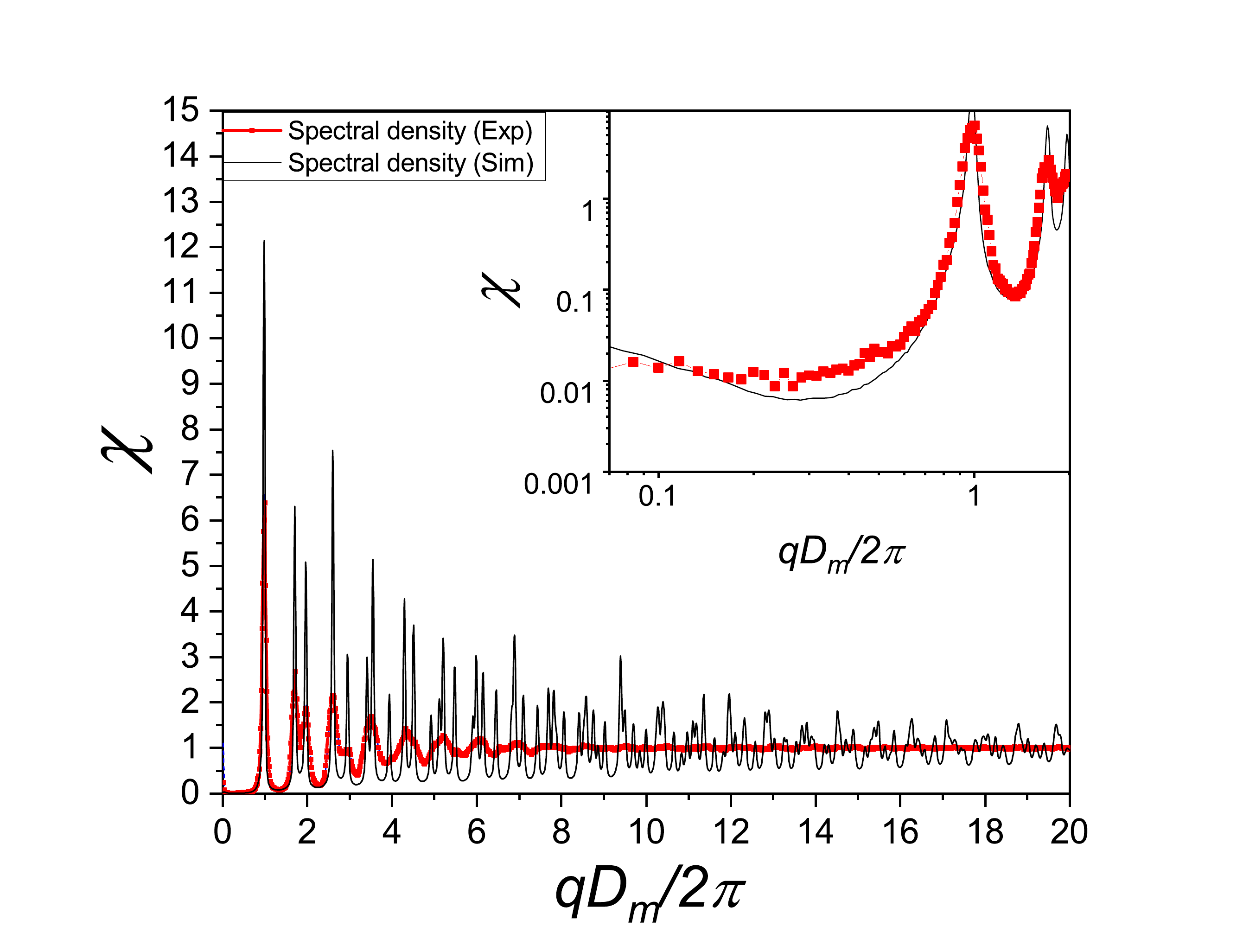}
\raisebox{7.3em}{$\textit{SR}=0.80$}\raisebox{6em}{\hspace{-4.55em}$\textit{NF}=0.62$}\; \raisebox{2em}{\includegraphics[width=0.2\linewidth]{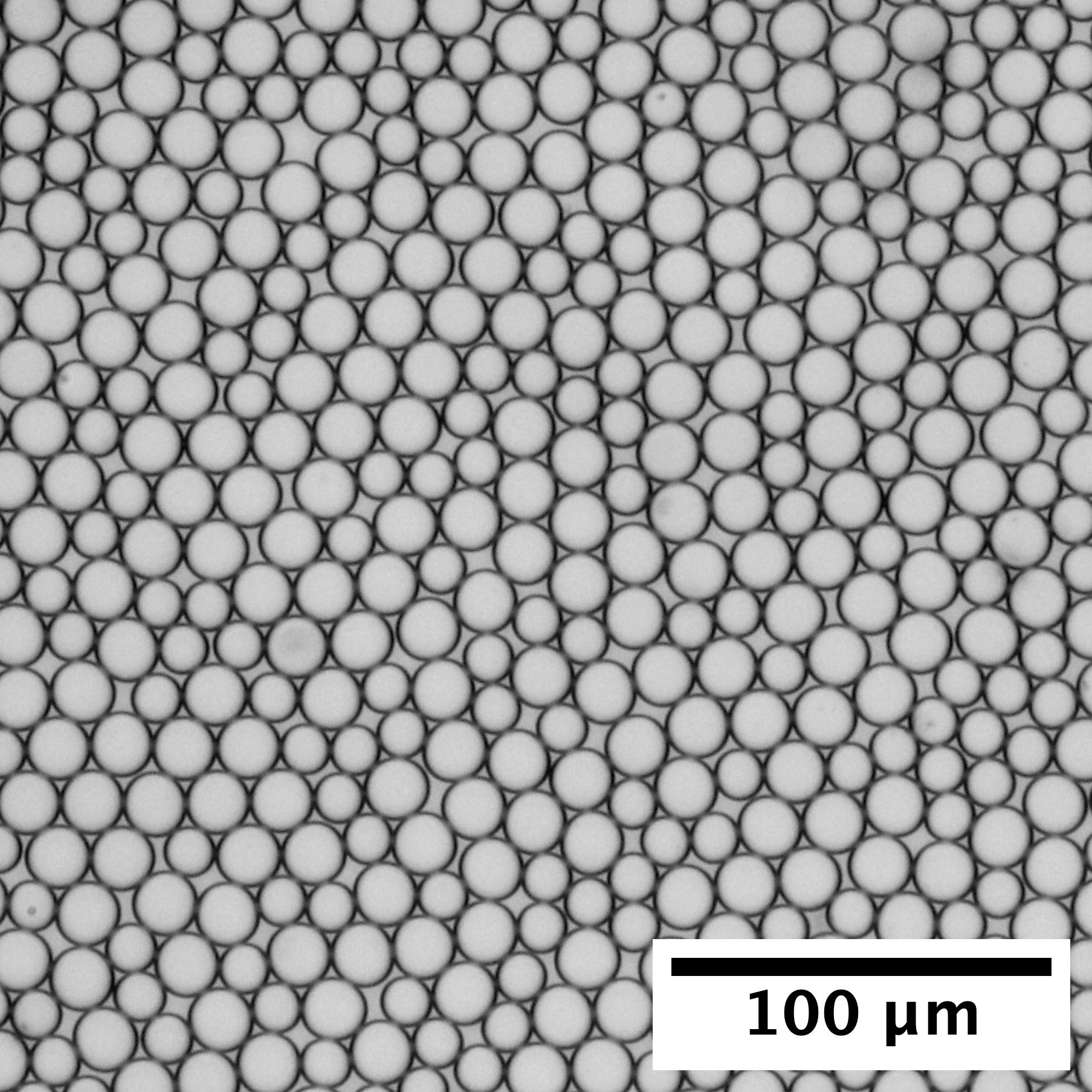}}\includegraphics[width=0.36\linewidth]{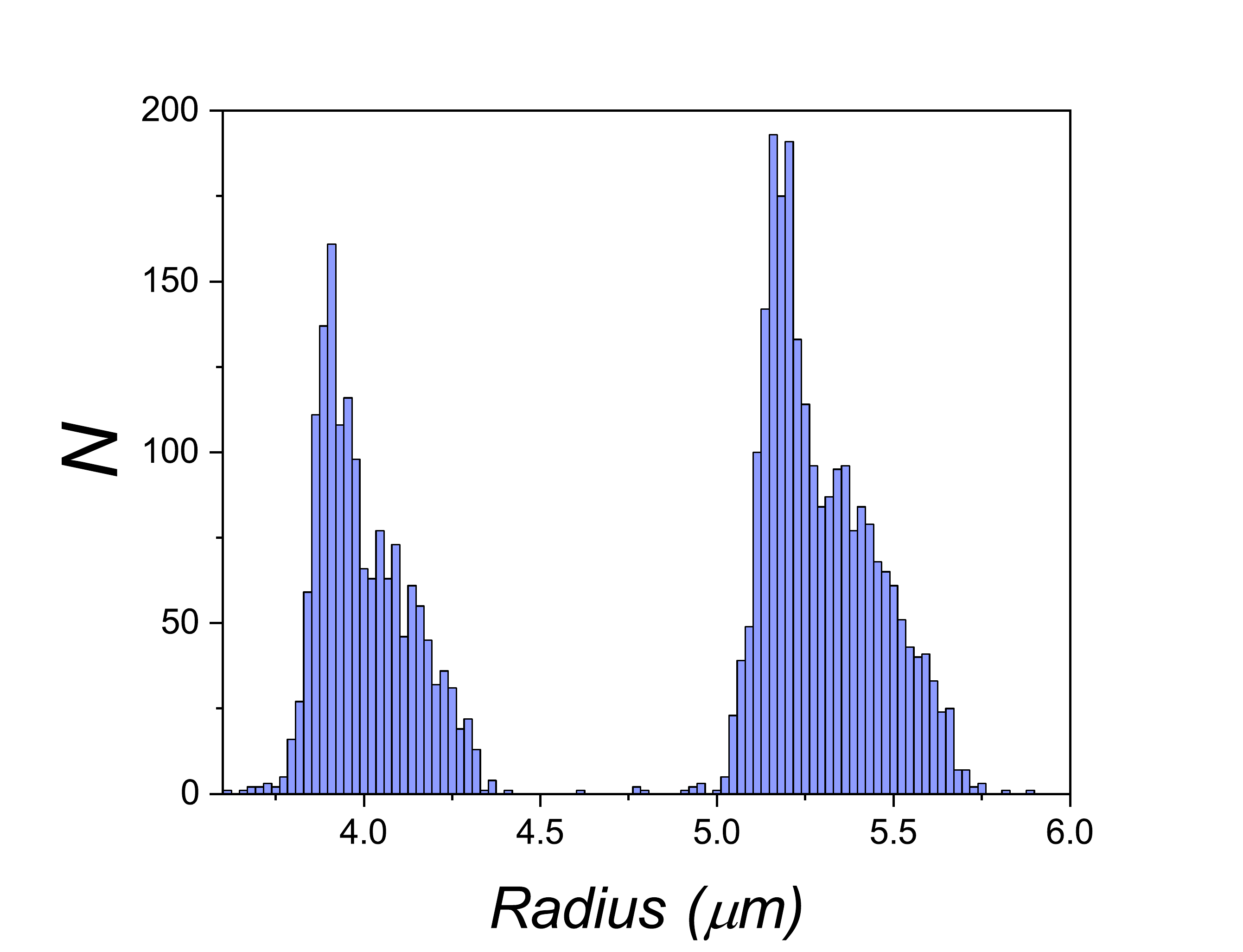}\hspace{-2.6em}\includegraphics[width=0.36\linewidth]{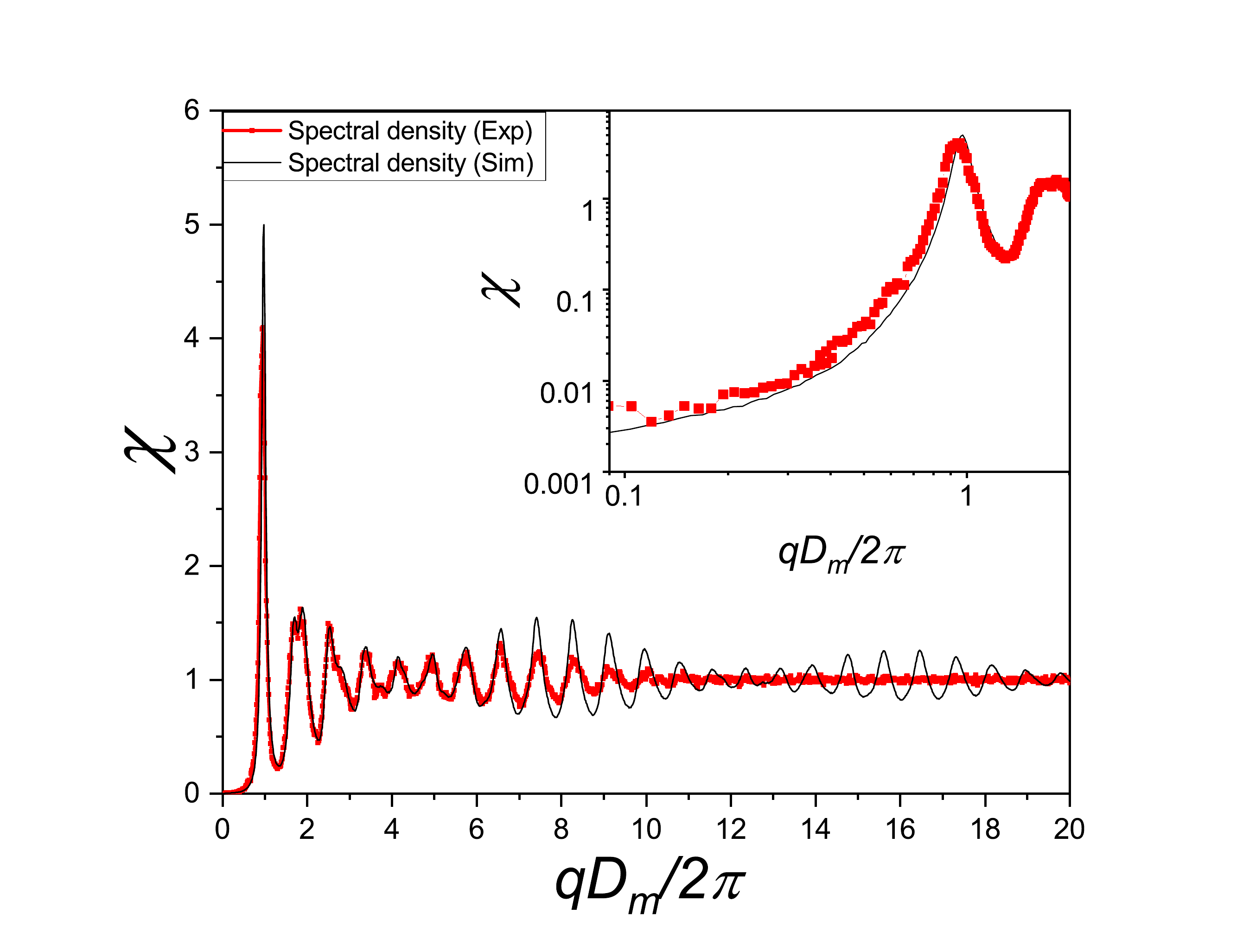}
\raisebox{7.3em}{$\textit{SR}=0.58$}\raisebox{6em}{\hspace{-4.55em}$\textit{NF}=0.50$}\; \raisebox{2em}{\includegraphics[width=0.2\linewidth]{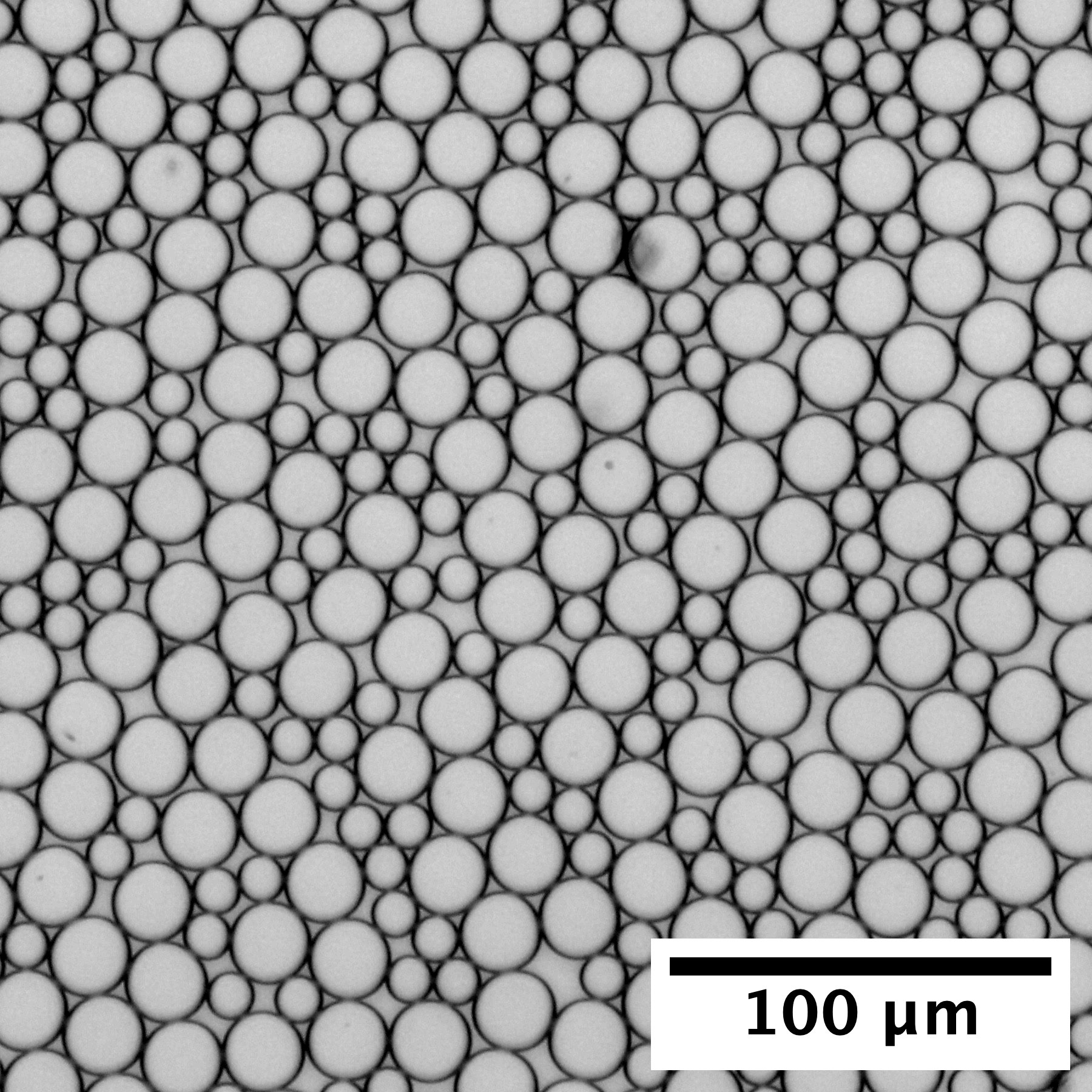}}\includegraphics[width=0.36\linewidth]{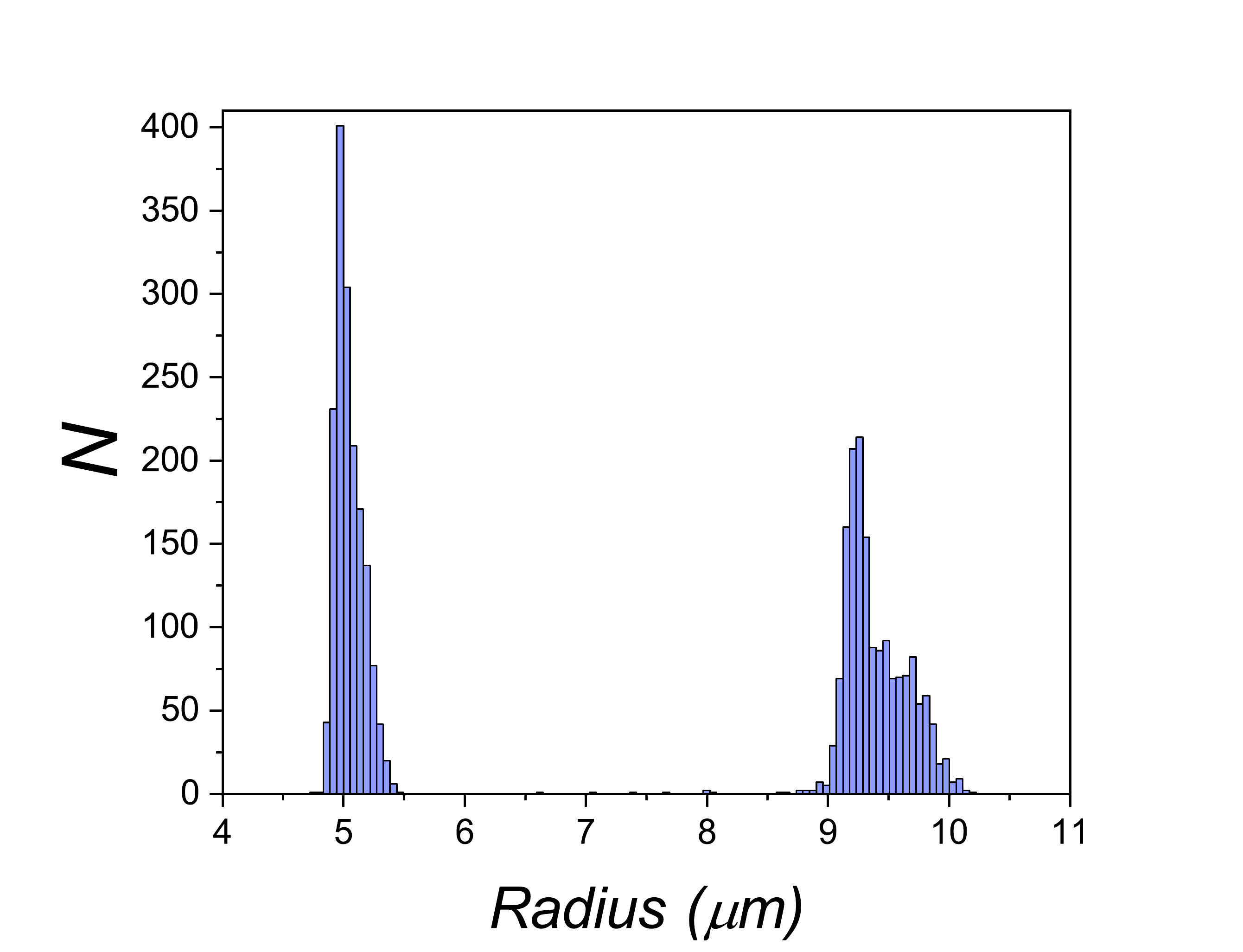}\hspace{-2.6em}\includegraphics[width=0.36\linewidth]{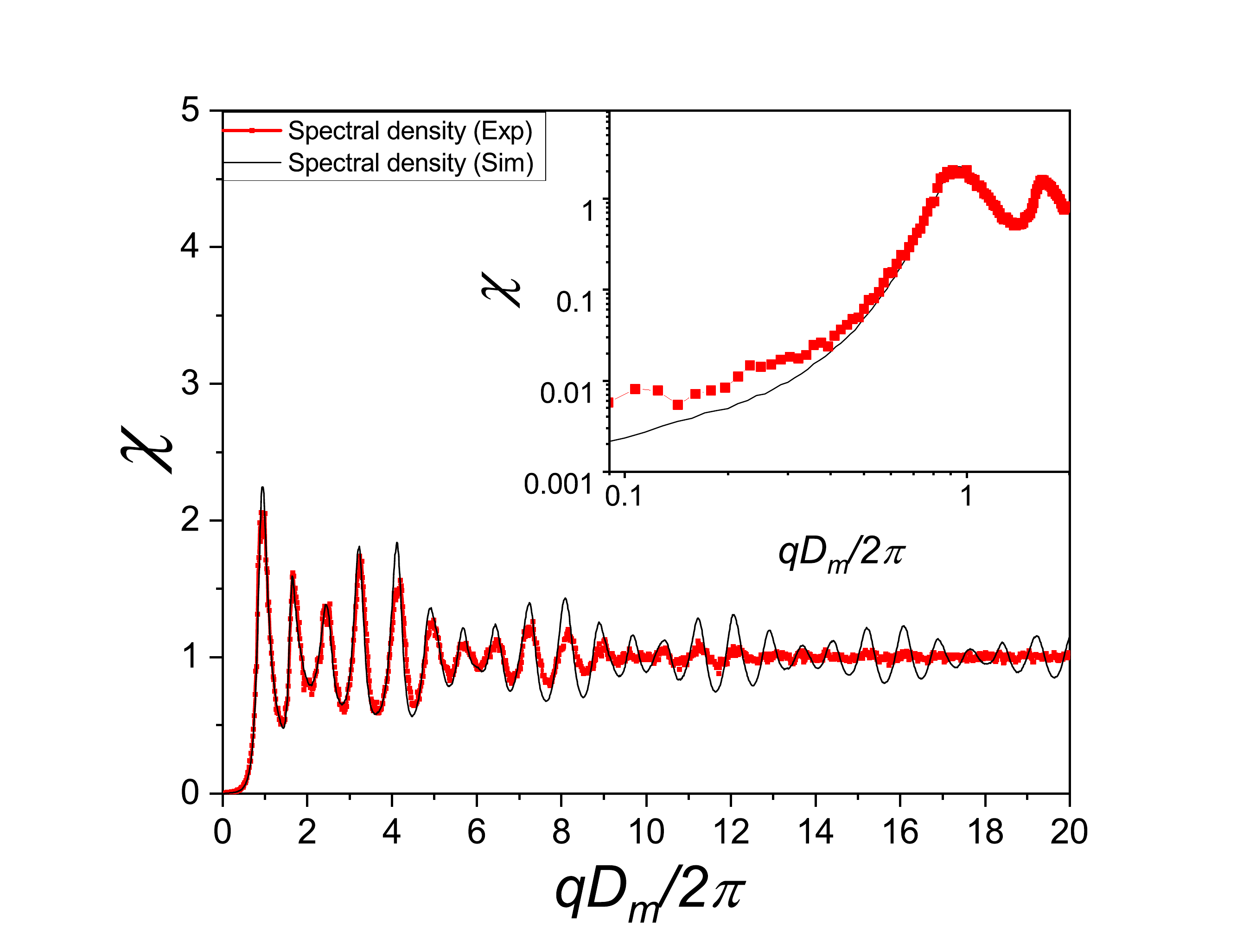}
\caption{Bidisperse mixtures. Left: Experimental closeup snapshots of the quasi-2D emulsion for several size ratios $\textit{SR}$ and number fractions $\textit{NF}$, as indicated. The values of $\textit{SR}$ and  $\textit{NF}$ are changed using different pressures at the inlet of the microfluidic chip. Center: Distribution of droplet radii, demonstrating approximate bidispersity. Right: Spectral densities as obtained from experiment (red connected dots) and  simulation (black line). The results from experiment and simulation match qualitatively, the oscillations in the experiment being more damped. Insets: Closeup on the spectral densities at small $q$, indicating the presence or absence of hyperuniformity. Data adapted from Ref.~\cite{ricouvier2017optimizing}.}
\label{fig:snaps+dist+chi}
\end{figure}

Figure~\ref{fig:snaps+dist+chi} shows three typical bidisperse mixtures. The top panels correspond to $\textit{SR}=0.80$ and $\textit{NF}=0.94$, such that most droplets are big. For this high $\textit{NF}$, the droplets form a polycrystalline structure. The crystalline domains are clearly visible, outlined by line defects which host the small droplets. Some small droplets can also be found within the crystalline lattice. The size distribution shows two peaks corresponding to the two populations of droplets. The standard deviation for each population is smaller than $0.05$, justifying the bidisperse approximation. The figure also shows the spectral density,
\begin{equation}
  \chi(\vecq) = \frac{1}{\overline{a^2}} \left\langle
  \frac{1}{N} \sum_{n,m} e^{-i\vecq\cdot\vecr_{nm}} a_n a_m \right\rangle,
\label{eq:chi}
\end{equation}
where $a_n$ is the area of droplet $n$, and $\overline{a^2}$ is the second moment of the droplet area distribution. In the terminology of Sec.~\ref{sec_gen}, $\chi(\vecq)$ is the CF with $\vecf(a)=a$; see Table~\ref{table:corr}. The spectral density shows damped Bragg peaks which are characteristic of a 2D hexagonal polycrystal. The first peak corresponds to the most common inter-droplet distance, denoted $D_m$. 

The middle panels of Fig.~\ref{fig:snaps+dist+chi} show the results for $\textit{SR} = 0.8$ and $\textit{NF} = 0.62$. This mixture is disordered and does not show extended crystal-like domains. The size distribution shows once again approximate bidispersity. The spectral density is more damped than for $\textit{NF}=0.94$ and does not exhibit the typical Bragg peaks of hexagonal lattices. At small $q$ it tends (linearly) toward zero, which is a signature of hyperuniformity~\cite{ricouvier2017optimizing,TorquatoBook}. Qualitatively similar behavior is seen for $\textit{SR} = 0.58$ and $\textit{NF} = 0.5$ in the bottom panels of Fig.~\ref{fig:snaps+dist+chi}. 

The spectral density obtained in the experiment is more damped than in the simulations. This is due to residual polydispersity and small deformations present in the experimental mixture of droplets, which randomize the correlations for large $q$, whereas the simulations are carried out for strictly bidisperse, nondeformable hard disks. The stronger disorder in the experimental system will translate below into a positive offset in the excess entropy.

The microscopic DOFs in the experiments and simulations are the locations of droplet centers, $\{\vecr_n\}=\{x_n,y_n\}$, and their areas, $\{a_n\}$. The two-point CFs that we consider are the following. (a) The structure factor, $S(\vecq)$, defined with $\vecf(a)=1$ in Eqs.~\eqref{eq:posfields}. (b) The spectral density, $\chi(\vecq)$, with $\vecf(a)=a$, which takes into account also the areas of the droplets~\cite{percus1958analysis,Book:Hansen}; see Eq.~(\ref{eq:chi}). (c) We consider also the mixed CF, $\vecM(\vecq)$, defined with $\vecf(a)=(1,a)$, imposing $S(\vecq)$ and $\chi(\vecq)$ simultaneously along with the position-area cross-correlations. See the SM~\cite{SM} for a detailed description of how these CFs are constructed and the way the excess entropies are calculated.

\begin{figure}
\includegraphics[width=0.5\linewidth]{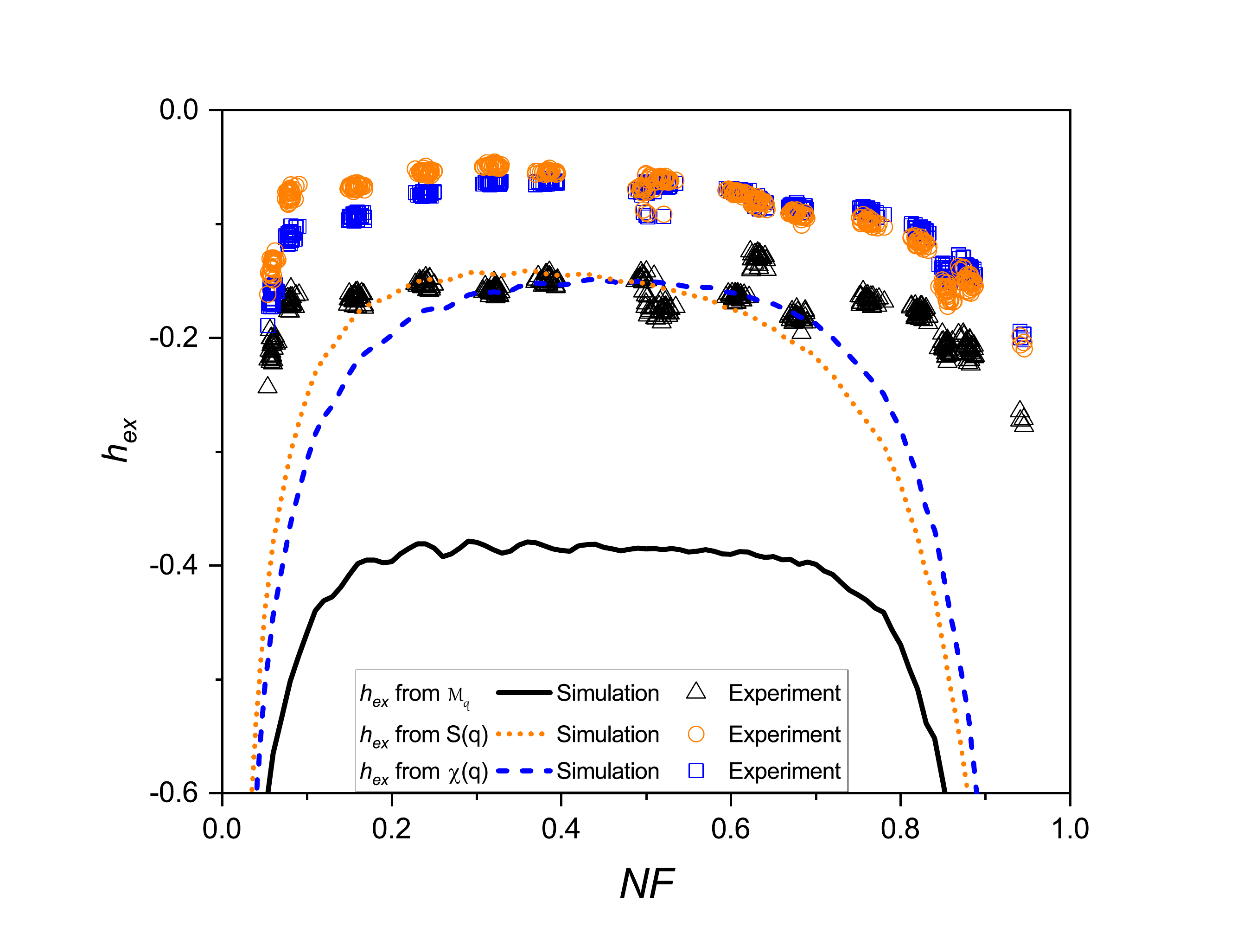}\hspace{-3.7em}\includegraphics[width=0.5\linewidth]{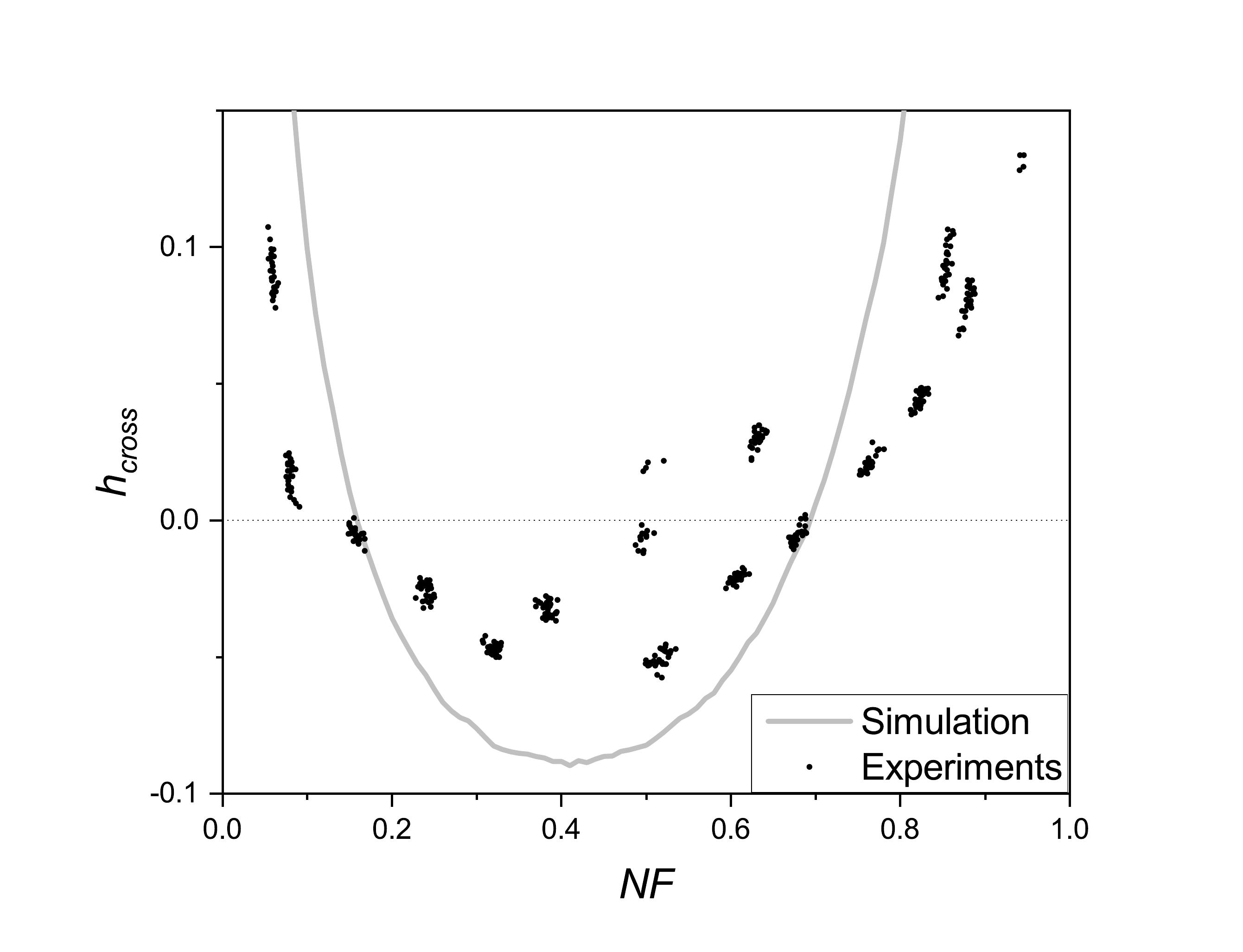}
\includegraphics[width=0.5\linewidth]{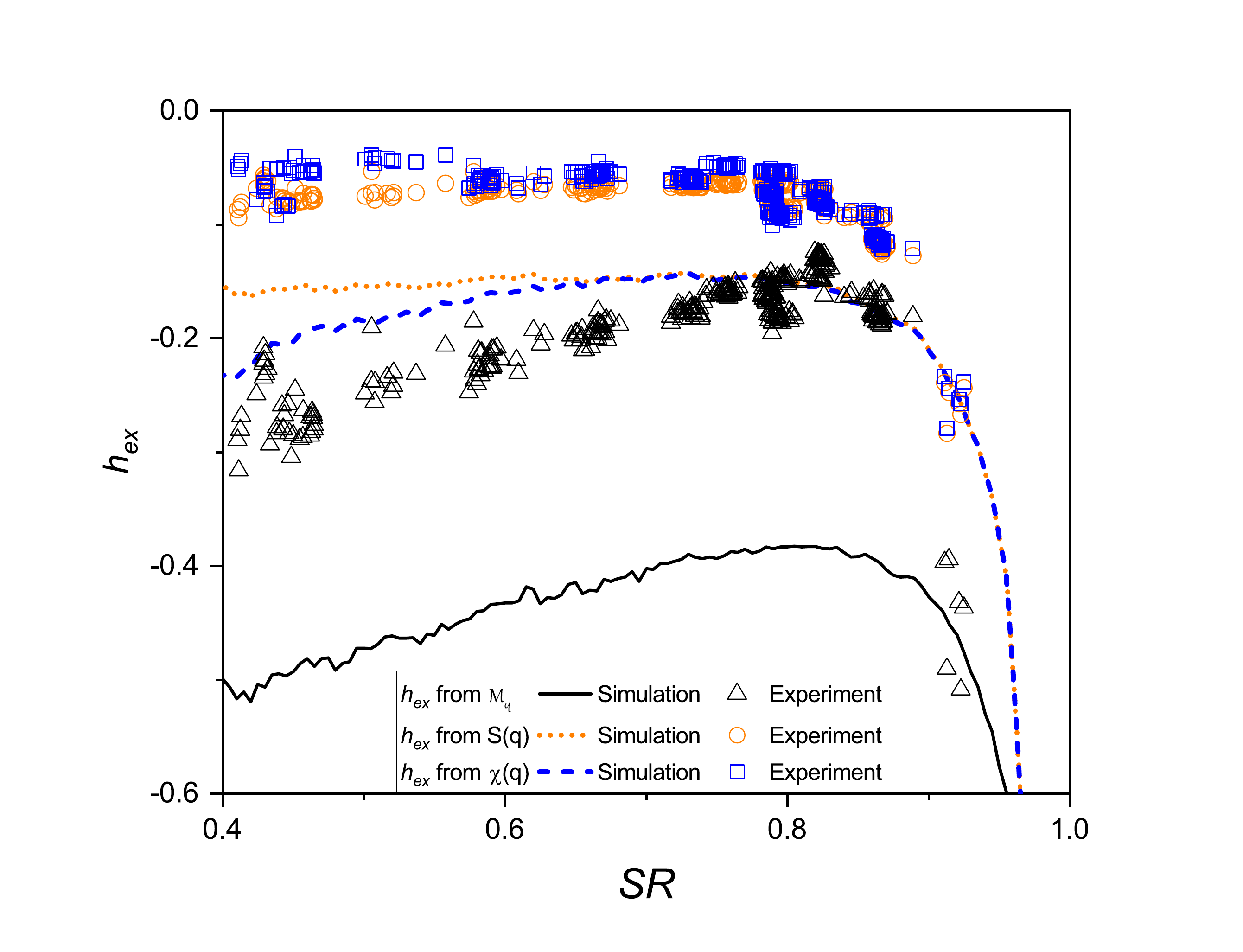}\hspace{-3.7em}\includegraphics[width=0.5\linewidth]{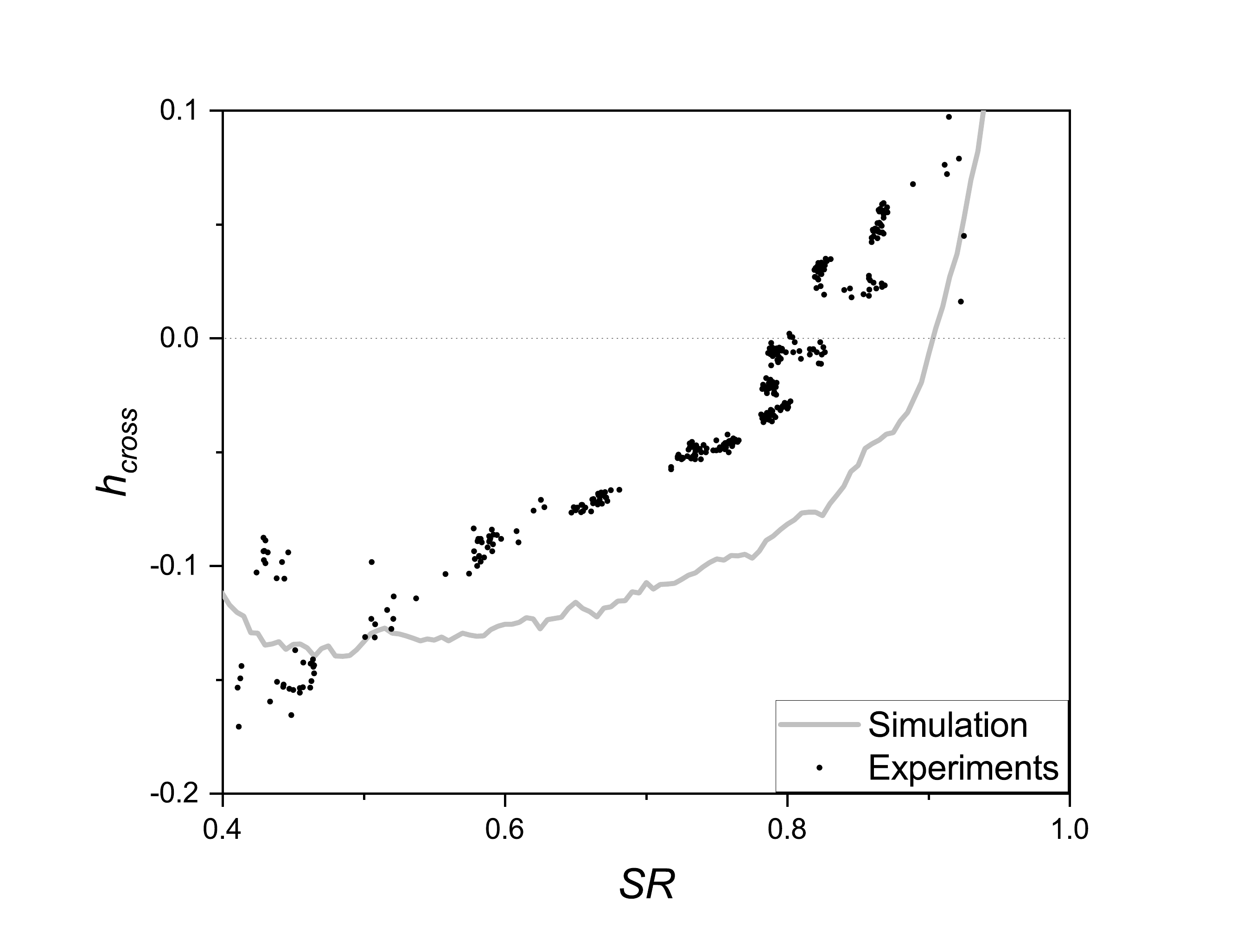}
\caption{Bidisperse mixtures. Entropy bounds as a function of number fraction $\textit{NF}$ (top panels, for $\textit{SR}\simeq0.8$) and size ratio $\textit{SR}$ (bottom panels, for $\textit{NF}\simeq0.5$). Left: Excess entropy per particle, $h_\mathrm{ex}$, as obtained from the structure factor $S_\vecq$ (experiment in orange {\color{orange} $\bigcirc$} and simulation in dotted orange line), spectral density $\chi_\vecq$ (blue {\color{blue} $\square$} and dashed blue line), and mixed correlation function $\vecM_\vecq$ (black $\triangle$ and black solid line). Right: Net entropy contribution of  position-area cross-correlations, $h_\mathrm{cross}$ (experiment in dots, simulations in solid line). The crossover from positive to negative values relates to the crossover between polycrystalline and disordered hyperuniform states.}
\label{fig:chimin+h(NF)}
\end{figure}

The excess-entropy bounds $h_\mathrm{ex}$, obtained from the measured CFs using Eq.~(\ref{eq:dischom}), are shown in the left panels of Fig.~\ref{fig:chimin+h(NF)}. Data points and lines correspond to results from  experiments and simulations, respectively. To demonstrate the robustness of the method, we have not ensemble-averaged the CFs; hence the many data points for each parameters set, showing a small spread. Note that by using $h_{\rm ex}$ (the excess entropy per particle relative to the uncorrelated system), the trivial dependencies of the entropy on the number of particles and field of view are removed. Note also that the field of view may contain different numbers of droplets (see, e.g., the snapshots in Fig.~\ref{fig:snaps+dist+chi}), which highlights the ability of the formalism to treat open systems. 

All three bounds in the left panels of Fig.~\ref{fig:chimin+h(NF)} show the expected simple trend\,---\,as the system gets away from the ordered crystalline state (i.e., away from $\textit{NF}=0,1$ in the top panel, and $\textit{SR}=1$ in the bottom one), the entropy increases. 
In the most ordered state, most droplets have six neighboring droplets, with a sixfold symmetry. Thus positions and cluster orientations are strongly correlated. When $\textit{SR}$ decreases or $\textit{NF}$ gets further away from $0$ or $1$, more possibilities open up for the immediate environment of a droplet. Therefore, the local orientation and the number of neighbors become less correlated, which increases the excess entropy.
The results from experiments and simulations agree qualitatively. As explained above, additional disorder in the experimental system leads to a positive entropy difference with respect to the simulation. While the distinction between the two phases is seen clearly also in the CF (the spectral density) of Fig.~\ref{fig:snaps+dist+chi}, it is less evident for close values of $\textit{NF}$s and $\textit{SR}$s (Figs.~S1 and S2~\cite{SM}). The entropy $h_\mathrm{ex}$, however, and especially  $h_\mathrm{cross}$, sharply  change in the crossover regions, indicating much more clearly the trends described above. For instance, for $\textit{SR}>0.76$ (where a droplet may have six neighbors; see Figs.~S1 and~S2), while the spectral density does not seem to change much (the Bragg peaks still persist), $h_\mathrm{cross}$ shows a significant drop, suggesting that the structural information does exist in the CF, but is subtly buried.

The entropy bounds nicely capture the asymmetry on the two sides of the symmetric composition $\textit{NF}=0.5$, i.e., $\textit{NF}$ and $(1-\textit{NF})$ are not statistically equivalent (top-left panel of Fig.~\ref{fig:chimin+h(NF)}). For example, the measured entropy for $\textit{NF}=0.2$ is higher than the one for $\textit{NF}=0.8$. Indeed, there are more possibilities to disperse small particles in a polycrystalline arrangement of big particles than the other way around. The asymmetry is most prominent for the $S(\vecq)$ bound, which is ``blind'' to the difference in particle sizes and reflects the difference in the number of positional arrangements as if the difference originated from some underlying interaction. This asymmetric behavior on the two sides of $\textit{NF}=0.5$ is a feature concealed in subtle differences between the CFs for different $\textit{NF}$s (see SM~\cite{SM}). In contrast, the asymmetry is clearly revealed by the entropy functional. 

The mixed CF $\vecM(\vecq)$ gives the lowest bound in both left panels. It constrains the structure factor and the spectral density simultaneously (along with cross-correlations), thus containing more information than each of the other two CFs. The bounds associated with $S(\vecq)$ and $\chi(\vecq)$ cross around $\textit{NF}\simeq 0.5$ in both experiment and simulation (top-left panel). This is because the spectral density is a ``re-weighting" of the structure factor in favor of the big particles (see Eq.~(\ref{eq:chi})). For small $\textit{NF}$ (majority of small droplets) this enhances the correlations as characterized by $\chi(\vecq)$, leading to a lower entropy, and the opposite happens for large $\textit{NF}$ (majority of large droplets). Comparing the bounds from $S(\vecq)$ and $\chi(\vecq)$ in the bottom-left panel, we find that the $S(\vecq)$ one is mostly higher in the simulation but slightly lower (better) in the experiment. This, too, may be a result of the residual polydispersity in the experiment, which weakens the correlations as measured by $\chi(\vecq)$.

The right panels of Fig.~\ref{fig:chimin+h(NF)} show the net entropy contribution of the position-area cross-correlations, defined as $h_\mathrm{cross}\equiv h_\mathrm{ex}[\vecM(\vecq)]-h_\mathrm{ex}[S(\vecq)]-h_\mathrm{ex}[\chi(\vecq)]$. In the absence of cross-correlations, it is easy to see from Eq.~\eqref{eq:dischom} that $h_\mathrm{ex}[\vecM(\vecq)]$ is equal to $h_\mathrm{ex}[S(\vecq)]+h_\mathrm{ex}[\chi(\vecq)]$, leading to $h_\mathrm{cross}=0$. Thus $h_\mathrm{cross}$ quantifies the additional information stored in the cross-correlations. However, $h_{\rm cross}$ as defined above may be negative or positive~\footnote{For orthogonal bases $\underline{f}$ (\ie $\vecA$ of Eq.~\eqref{eq:norm} is diagonal), $h_{\rm cross}\leq0$ always, which is not the case here. For example, $h_{\rm cross}$ from a CF that mixes position and orientation would be non-positive, since the basis is a subset of circular/spherical harmonics.}. This is because $S(\vecq)$ and $\chi(\vecq)$ are CFs of DOFs which are not independent (both CFs include position correlations). For example, in the limit of a monodisperse system, $\chi(\vecq)$ and $S(\vecq)$ coincide, and we have $h_{\rm ex}[\vecM(\vecq)]=h_{\rm ex}[S(\vecq)]=h_{\rm ex}[\chi(\vecq)]$, leading to $h_{\rm cross}=-h_{\rm ex}[S(\vecq)] > 0$. See Fig.~\ref{fig:chimin+h(NF)} at small and large $\textit{NF}$, and large $\textit{SR}$. Position-area cross-correlations become strong when the spatial arrangement of nearby droplets is affected by their sizes. For instance, a droplet can be surrounded by six others when $\textit{SR}$ is close to 1, and by seven smaller ones if $\textit{SR}<0.76$. This accounts for the sharp decrease of $h_{\rm cross}$ with decreasing $\textit{SR}$ (bottom-right panel).

Overall, the key observation in both right panels of Fig.~\ref{fig:chimin+h(NF)} is the dominant contribution of the cross-correlations between position and area to the entropy of the heterogeneous, disordered and hyperuniform system.

\section{Discussion}
\label{sec_discussion}


Any measurement provides information on the system of interest and thus lowers the upper bound on its entropy.  Equations~\eqref{eq:dischom}, \eqref{eq:cont_hom_ent}, and~\eqref{eq:hent} quantify this entropy reduction quite generally when the measurement is of a pair-correlation function. The extent to which these results are useful has been examined in Sec.~\ref{sec_experiment}, where we have used the entropy bounds to investigate subtle features of a transition out of equilibrium.

The central advantage of the approach presented here is the ability to not only estimate the entropy but also separate it into distinct contributions coming from different DOFs and their inter-correlations. This ``finer resolution" intimately connects the method to the physics of the system under study. When applied to jammed  mixtures (Sec.~\ref{sec_experiment}), the entropy bounds have revealed the crucial role played by cross-correlations between the particles' positions and sizes. Previous works associated the formation of disordered hyperuniform structures solely with strong positional correlations~\cite{torquato:review}. Our results suggest that, in polydisperse systems, the entropy reduction due to hyperuniformity is dominated by strong position-size cross-correlations.

Another key feature of the formalism is that, by choosing measurable, concrete constraints (\ie CFs), we essentially bypass the undersampling problem of entropy estimation, integrating out the unconstrained information. Importantly, any symmetry (or symmetry breaking) reflected in the CFs is by construction taken into account. The cost of these advantages is that the method is not a ``black box'' which  can be blindly applied to any data, but requires knowledge of  the system's DOFs.

A few important technical advantages should be noted as well. (a) The computational cost of the method does not scale with system size. (b) There is no artificial discretization of phase space. (c) Generalization to include other observables is achieved by including the corresponding constraints. This flexibility, for example, has allowed us to include the effects of varying particle number and external fields (see Appendix~\ref{appendix:inhom}). 

The formalism is restricted in two main respects. (a) It is limited to upper bounds for the entropy, which might not always be sufficiently tight to be useful. In such cases one may improve the bound, e.g., by including higher-order correlations (see, for example, the useful high-order correlations measured in Ref.~\cite{pnas2009x}.) (b) The main results, Eqs.~\eqref{eq:dischom}, \eqref{eq:cont_hom_ent}, and~\eqref{eq:hent}, involve a Gaussian approximation. One may improve them using standard perturbation techniques. Indeed, some constraints may give rise to equations which cannot be solved analytically, which will require numerical solutions~\cite{Zhang2020}.

The possibilities to apply entropy-bound functionals to additional systems are vast. We mention two examples which seem particularly appealing. Indications of entropy changes occurring at the glass transition~\cite{TruskettPRE2000,MittalJCP2006} may be checked in more detail, similar to the case of the jammed mixtures in Sec.~\ref{sec_experiment}. If such changes are found, one might be able to point at the dominant DOFs, or their interactions, underlying the transition. This particular application is likely to require the inclusion of higher-order correlations. Active matter exhibits dynamic transitions, such as bacterial swarming~\cite{be2020phase} and motility-induced phase separation~\cite{cates2015motility}. Resolving different contributions to the entropy may provide new insights into these far-from-equilibrium phenomena.

Finally, it may be possible, and highly desirable, to go beyond steady states and use a similar approach for time-dependent phenomena. Thus, for example, the contributions of different DOFs to the entropy production might be resolved by exploiting temporal CFs as dynamical constraints.

\begin{acknowledgments}
  We thank David Andelman, Roi Peer, Shlomi Reuveni, and Yael Roichman for helpful suggestions. 
  We thank Patrick Tabeling and Pavel Yazhgur for sharing the experimental data.
  H.D. and B.S. acknowledge support from the Israel Science Foundation (Grant No.\ 986/18). J.R. is grateful to the Azrieli Foundation for the award of an Azrieli Fellowship.
\end{acknowledgments}


\setcounter{section}{0}
\newlength{\pointwidth}
\settowidth{\pointwidth}{.}
\renewcommand{\thesection}{APPENDIX \Alph{section}}



\section{Entropy bound for inhomogeneous systems}
\label{appendix:inhom}

\renewcommand{\theequation}{A\arabic{equation}}
\setcounter{equation}{0}

The purpose of this Appendix is to extend the formalism described in Sec.~\ref{sec_gen} to systems whose translational symmetry is broken by some externally applied force. The adjusted bound is given in Eq.~\eqref{eq:hent} below. In addition, this Appendix will facilitate the detailed derivation of Eq.~\eqref{eq:dischom} in Appendix~\ref{appendix:gencor}.

\begin{subequations}
\label{Eq:2pcorr&profile}
As before, the key observable is a two-point CF. We rely on the setup presented in Sec.~\ref{sec_gen}. First, we define the (Fourier-transformed) field of interest following Eqs.~\eqref{eq:posfields}. Now, due to the inhomogeneity, two distinct $\vecq$ modes might have nonzero correlations. Hence, the CF of Eq.~\eqref{eq:homcorr} is extended to include off-diagonal terms in $\vecq$,
\begin{eqnarray}
    \bphi(\vecq,\vecq')=\vecA^{-1}\langle N^{-1}\vecJ(\vecq)\vecJ^\dagger(\vecq')\rangle,
    \label{eq:inhomcorr}
\end{eqnarray}
where $\vecA$ is the same normalization matrix defined in Eq.~\eqref{eq:norm}.

In addition to the more complex CF, the average field $\langle\vecJ(\vecq)\rangle$ may be nonuniform. For example, the steady-state density of particles (using $\vecf=1$) may be inhomogeneous due to gravity. We characterize these effects by introducing the average profile,
\begin{equation}
    \vecG(\vecq)=\langle N^{-1/2}\vecJ(\vecq)\rangle,\label{eq:prof}
\end{equation}
\end{subequations}
as another constraint. The CFs and profiles may be available from experiments (\eg scattering and absorption, respectively) and simulations. As before, we exclude the uniform $\vecq=\veco$ mode.

A few additional definitions are required. We recall that only a finite set $\{\vecq_1,\vecq_2,\ldots,\vecq_\Omega\}$ of $\vecq$ modes is considered. 
It is useful to consider $\vecq$ as an additional index in an extended matrix that treats the $\Omega$ modes $\vecq$ and the $M$ components of $\vecf$ together.
We define the $(\mq \mf)\times(\mq \mf)$ matrices $\hat{\bphi}$, $\hat{\vecA}$, and the unit matrix $\hat{\mathcal{I}}$, and similarly the $(\mq\mf)$-long vector $\hat{\vecG}$. These are just $\mq\times \mq$ blocks of the $M\times M$ matrices $\bphi (\vecq,\vecq')$, $\mathcal{I}$, and $\mathcal{\vecA}$, and $\mq$-times stacked $M$-long vector $\vecG (\vecq)$,
\begin{align}
    \hat{\bphi}=\left(\begin{array}{c|c|c|c}
\bphi(\vecq_1,\vecq_1) & \bphi(\vecq_1,\vecq_2) & \cdots & \bphi(\vecq_1,\vecq_\mq) \\ \hline
\bphi(\vecq_2,\vecq_1) & \bphi(\vecq_2,\vecq_2) & \cdots & \bphi(\vecq_2,\vecq_\mq) \\ \hline
\vdots & \vdots & \ddots & \vdots \\ \hline
\bphi(\vecq_\mq,\vecq_1) & \bphi(\vecq_\mq,\vecq_2) & \cdots & \bphi(\vecq_\mq,\vecq_\mq)
\end{array}\right),\;
    \hat{\vecA}^{-1}=\left(\begin{array}{c|c|c|c}
\vecA^{-1} & \calo & \cdots & \calo \\ \hline
\calo & \vecA^{-1} & \cdots & \calo \\ \hline
\vdots & \vdots & \ddots & \vdots \\ \hline
\calo & \calo & \cdots & \vecA^{-1}
\end{array}\right),\nonumber\\
    \hat{\vecG}=\left(\begin{array}{c|c|c|c}
\vecG^T(\vecq_1) & \vecG^T(\vecq_2) & \cdots & \vecG^T(\vecq_\mq)
\end{array}\right)^T.\label{eq:megaM}
\end{align}

Now that the constraints have been defined and rearranged, the entropy bound obtained for a two-point CF $\bphi(\vecq,\vecq')$ and a steady profile $\vecG(\vecq)$ can be written as,
\begin{equation}
    h_{\rm ex}[\bphi,\vecG] = \frac{1}{2\Nmean} \tr_{\mq\mf}\left[\ln_{\mq\mf}(\hat{\bphi}-\hat{\vecA}^{-1}\hat{\vecG}\hat{\vecG}^\dagger)+\hat{\mathcal{I}}-\hat{\bphi}\right].
    \label{eq:hent}
\end{equation}
The detailed derivation is given in the SM~\cite{SM}. The matrix logarithm and trace are applied to the objects of Eq.~\eqref{eq:megaM}, and the subscripts $\mq\mf$ are a reminder that they are trace and matrix-log operations performed on $(\mq\mf)\times(\mq\mf)$ matrices. The number of particles is allowed to vary, with a given average number $\Nmean$, and with a CF $\bphi(\vecq,\vecq')$ and steady profile $\vecG(\vecq)$ which take into account the varying $N$ through Eqs.~\eqref{Eq:2pcorr&profile}. 

We may interpret Eq.~\eqref{eq:hent} by considering two opposite limits. In systems of single-particle interactions (\ie in the presence of a strong external field), the two-point CF becomes unity, $\bphi(\vecq,\vecq')=\mathcal{I}\delta(\vecq-\vecq')$ (\ie $\hat{\bphi}=\hat{\mathcal{I}}$), and the bound reduces to 
\begin{equation}
    h_{\rm ex}[\vecG]=\frac{1}{2\Nmean} \tr\left[\ln(\hat{\mathcal{I}}-\hat{\vecA}^{-1}\hat{\vecG}\hat{\vecG}^\dagger)\right]\simeq-\frac{1}{2\Nmean}\sum_{\vecq\ne\veco}\vecG^\dagger(\vecq)\vecA^{-1}\vecG(\vecq).\label{eq:inhoment}
\end{equation}
This agrees to second order (due to the Gaussian approximation in the derivation) with the trivial result $h_{\rm ex}=-\int d\vecr d\vecj \prob(\vecr,\vecj)\ln (\upsilon \prob(\vecr,\vecj))$, where $\prob(\vecr,\vecj)$ is the joint single-particle distribution of $\vecr$ and $\vecj$, and $\upsilon=\int d\vecr d\vecj$.
In the other limit of homogeneous systems, the steady profile is absent, $\vecG=\bbo$, and the CF is $\vecq$ diagonal (as in Eq.~\eqref{eq:homcorr}). The expression for the entropy bound reduces to Eq.~\eqref{eq:dischom},
\begin{eqnarray}
    h_{\rm ex}[\bphi]&=&\frac{1}{2\langle N\rangle} \sum_{\vecq\ne\veco}\tr\left[\ln\bphi(\vecq,\vecq)+\mathcal{I}-\bphi(\vecq,\vecq)\right].\nonumber
\end{eqnarray}
Here, the matrix logarithm and trace are of $\mf\times\mf$ matrices only, and the summation over $\vecq$ is explicitly written. Comparing to Eq.~\eqref{eq:hent}, we see that, in the absence of a steady profile ($\vecG=\bbo$), the only component in the field's ``variance',' $\hat{\bphi}-\hat{\vecA}^{-1}\hat{\vecG}\hat{\vecG}^\dagger$, that survives is the second-moment matrix (its $\vecq$ diagonal terms), $\bphi(\vecq,\vecq)$. Together, Eqs.~\eqref{eq:inhoment} and~\eqref{eq:dischom} demonstrate how Eq.~\eqref{eq:hent} could be separated into single-particle and two-particle contributions. Also, they do not require the more cumbersome and computationally costly objects of Eq.~\eqref{eq:megaM}.

\section{Derivation of the central result}
\label{appendix:gencor}

\renewcommand{\theequation}{B\arabic{equation}}
\setcounter{equation}{0}

Here we give a detailed derivation of the main result\,---\,an upper bound for the entropy given a general two-point CF. A steady-state average field, and a grad-canonical ensemble are considered in the SM~\cite{SM}.

Following the main text, we deal with systems whose particles possess more DOFs than just locations. A complete microscopic configuration is therefore described by $\{\vecx_n\}=\{\vecr_n,\vecj_n\}$, where $\vecr_n$ is the spatial position of the $n$th particle and the vector $\vecj_n$ lists the scalar components of its additional DOFs. 

We consider an arbitrary vectorial function of the DOF $\vecj$, to obtain a field as written in Eq.~\eqref{eq:realf}, $\vecJ(\vecr)=\sum_n\delta(\vecr-\vecr_n) \vecf(\vecj_n)$, where $\vecf \in \mathbb{C}^M$ is a column vector. A simple example is the case of 3Dorientations, $\vecj=(\vartheta,\varphi)$, where the function is the decomposition into spherical harmonics, arranged in a column vector $\vecf_{l^2+(m+l)+1}(\vecj)=Y_l^m(\vartheta,\varphi)$. Next, a Fourier transform is taken to obtain Eq.~\eqref{eq:q-f}, $\vecJ(\vecq)=\sum_ne^{-i\vecq\cdot\vecr_n}\vecf(\vecj_n)$. For shorthand, we replace $e^{-i\vecq\cdot\vecr_n}\vecf(\vecj_n)$ with a simple $\hat{\vecg}(\vecx_n)$ (the discrete $\vecq$ absorbed as an index, as in the objects of Eq.~\eqref{eq:megaM}), such that the transformed field is
\begin{equation}
    \hat{\vecJ}=\sum_n\hat{\vecg}(\vecx_n).
\end{equation}
Proceeding with this notation, where the position DOF is not written explicitly, highlights the fact that the formalism (and the bound appearing in Eq.~\eqref{eq:hent}) applies for nonspatial CFs as well.

The two constraints to be imposed are the following. (a) The CF (Eq.~\eqref{eq:inhomcorr}),
\begin{subequations}
\begin{equation}
    \hat{\bphi}=\langle \hat{\vecB}^{-1}\hat{\vecJ}\hat{\vecJ}^\dagger\rangle,\label{eq:matcorr}
\end{equation}
where $\hat{\vecB}\equiv N\hat{\vecA}=N\int d\vecx \prob(\vecx)\hat{\vecg}(\vecx)\hat{\vecg}^\dagger(\vecx)$ is a the normalization matrix, and $\prob(\vecx)$ the single-particle distribution. (b) If the system is inhomogeneous, we additionally constrain the steady-state profile (Eq.~\eqref{eq:prof}),
\begin{equation}
    \hat{\vecG}=\langle N^{-1/2}\hat{\vecJ}\rangle.
\end{equation}
\end{subequations}
Recall that we exclude the uniform mode (for the spatial field, the $\vecq=\veco$ mode); hence, $\vecg$ is not a complete basis (as one component is missing). This will come into play in the calculation performed below (at the saddle-point approximation step).

As mentioned in the main text, experimentally, a finite number ($\mq$) of $\vecq$ modes are considered, which only loosens the entropy bound. Similarly, including only a finite number ($\mf$) of the transform's components, $\vecf$, would merely loosen the bound, too. In the examples of 3D orientations, taking $\vecf=\hat{\vecn}$ implies the inclusion of only the $M=3$ terms $Y_1^{0,\pm1}$, and taking $\vecf$ as the independent components of the nematic tensor implies the inclusion of the $M=6$ terms $Y_0^0,Y_2^{0,\pm1,\pm2}$. Thus, as established in Appendix~\ref{appendix:inhom}, $\hat{\vecG},\hat{\vecJ},\hat{\vecg}(\vecx_n)\in\mathbb{C}^{\mf\mq}$, and $\hat{\bphi}$, $\hat{\vecB}$, and $\hat{\mathcal{I}}$ are $\mf\mq\times\mf\mq$ matrices. While $\mf\mq\equiv\mt$ is finite in practice, for the sake of the derivation we assume that $\hat{\vecg}(\vecx)$ (up to adding the uniform mode) is a complete basis for $\vecx$. (This subtlety is required because later in the derivation we rely on the existence of the inverse transform.) We will drop the hats of the $\mt$-sized column vectors and $\mt\times\mt$ matrices; the ``hatless" $\mf$-sized column vectors and $\mf\times\mf$ matrices (shown in Sec.~\ref{sec_gen} and Appendix~\ref{appendix:inhom}) do not appear here. The connection to actual physical observables is more evident from the notation adopted in the main text. The more elaborate notation used here and in the SM~\cite{SM} makes the detailed derivation more concise.

We maximize the entropy (Eqs.~\eqref{eq:shan}) over the probability density function $p(\vecX)$ of microstates $\vecX=\{\vecx_n\}$, under two constraints: (a) normalization, $\int \dX p(\vecX)=1$, and (b) the key constraint\,---\,a given CF (Eq.~\eqref{eq:matcorr}),
\begin{equation}
    \bphi=\vecB^{-1}\int \dX p(\vecX)\sum_{n=1}^N\sum_{m=1}^N\vecg(\vecx_n)\vecg^\dagger(\vecx_m).\label{eq:MegaCF}
\end{equation}
We denote $\dX=d\vecX\prod_{n=1}^N\prob(\vecx_n)$ to allow the case of quenched DOFs, whereby the single-particle distribution $\prob(\vecx)$ serves as a ``weight function''. Similarly, $\dx=d\vecx\prob(\vecx)$. In this case $p(\vecX)$ may be interpreted either as the $N$-particle distribution function~\cite{Book:Hansen}, or, up to a change of variables ($\vecx \to\mathrm{CDF}(\vecx)$, the cumulative distribution function), the copula density~\cite{Book:copula}. Therefore, $p(\vecX)$ is related to the probability density function to find the system in microstate $\vecX$, $P(\vecX)$ (the one appearing in Eq.~\eqref{eq:cont_shan}), through $P(\vecX)=p(\vecX)\prod_{n=1}^N\prob(\vecx_n)$. In the simpler case of annealed DOFs, $\prob(\vecx)=1/\upsilon$, where $\upsilon=\int d\vecx$ is the phase-space volume.

The thermodynamic entropy, following Eqs.~\eqref{eq:shan}, is
\begin{eqnarray}
    S&=&-\int d\vecX \prod_{n=1}^N\prob(\vecx_n)p(\vecX)\ln\left[\prod_{n=1}^N\prob(\vecx_n)p(\vecX)\right]\nonumber\\
    &=&-\int d\vecX \prod_{n=1}^N\prob(\vecx_n) p(\vecX) \ln p(\vecX)+Ns_1,\label{eq:alt_ent}
\end{eqnarray}
where $s_1=-\int d\vecx \prob(\vecx)\ln\prob(\vecx)$ is the entropy of a single isolated particle (an {\it a priori} known constant). In the absence of inter-particle coupling, $N s_1$ is the only surviving contribution to the entropy; hence, we refer to it as the ideal gas contribution, $S^\mathrm{id}=Ns_1$, and define the excess entropy as
\begin{equation}
    S-S^\mathrm{id}=-\int \dX p(\vecX) \ln p(\vecX).\label{eq:exent}
\end{equation}

In order to find the bound $H_\mathrm{ex}\geq S-S^\mathrm{id}$, we seek to maximize the Lagrangian 
\begin{eqnarray}
  {\mathcal L} &=&-\int \dX p(\vecX) \ln p(\vecX)- \eta \left( \int \dX p(\vecX) - 1 \right) \nonumber\\
   &&- \tr\left[ \blam^\dagger \int \dX p(\vecX)\left(\sum_{n,m}\vecg(\vecx_n)\vecg^\dagger(\vecx_m) - \vecB \bphi \right) \right],
\label{eq:lagrangian}
\end{eqnarray}
such that $\bphi$ is the given CF, and find the entropy-maximizing $p(\vecX)$ which reproduces that particular $\bphi$. Here, $\blam$ (a $\mt\times\mt$ matrix) and $\eta$ (scalar) are the Lagrange multipliers imposing the constraints, and the trace is of $\mt\times\mt$ matrices. The trace notation is a convenient shorthand arising from the matrix identity $\tr \vecA^\dagger \vecB = \sum_{ii'} \vecA^*_{ii'} \vecB_{i'i}$ (the Frobenius matrix inner product). Equation~\eqref{eq:lagrangian} is equivalent to imposing every component of $\vecB\bphi$ as a separate constraint. We should solve the equations
\begin{equation}
   \frac{\delta {\mathcal L} }{\delta p(\vecX)}  = 0
   , \qquad \frac{\partial {\mathcal L} }{\partial \eta}=0 ,\qquad \frac{\partial {\mathcal L} }{\partial \blam^\dagger} = \calo, \label{eq:constra}
\end{equation}
where $(\partial/\partial \blam^\dagger)$ denotes $\mt^2$ partial derivatives with respect to each component of $\blam^\dagger$. 

Taking the variation with respect to $p(\vecX)$ yields,
\begin{equation}
    p(\vecX)=e^{-1-\eta} \Gamma(\vecX),\qquad \Gamma(\vecX)=\exp\left(-\tr\left[ \blam^\dagger \left(\sum_{n,m}\vecg(\vecx_n)\vecg^\dagger(\vecx_m) - \vecB \bphi \right) \right] \right).\label{eq:gamma}
\end{equation}
Substituting into the normalization condition for $p(\vecX)$, we find 
\begin{equation}
    \eta=\ln \z-1,\qquad \z = \int\dX\Gamma(\vecX),\label{eq:partition}
\end{equation}
leading to $p(\vecX)= \z^{-1} \Gamma(\vecX)$. We see that $\z$ and $\blam$ play the analogous roles of the partition function (relative to the ideal gas) and pair-interaction, inside a Boltzmann-like factor. This analogy is a result of the variational procedure, despite the nonequilibrium setting of the formalism. The main difference from conventional equilibrium calculations is that we do not know the effective pair interaction, $\blam$, which is yet to be found.

Next, consider the derivative of the partition function with respect to the components of $\blam^\dagger$, 
\begin{equation}
    \frac{\partial\ln\z}{\partial\blam^\dagger}=\z^{-1}\frac{\partial}{\partial\blam^\dagger}\int \dX \Gamma(\vecX)=\int \dX p(\vecX)\sum_{n,m}\vecg(\vecx_n)\vecg^\dagger(\vecx_m) - \vecB \bphi=\calo.\label{eq:enforceZ}
\end{equation}
Therefore, solving $\partial\ln\z / \partial\blam^\dagger=\calo$ is equivalent to forcing the constraints on the CF. The maximum entropy is obtained as
\begin{equation}
    H_\mathrm{ex}=-\int \dX p(\vecX) \ln p(\vecX)=\int \dX p(\vecX) \ln\z-\int \dX p(\vecX) \ln \Gamma(\vecX)=\ln\z,\label{eq:entpart}
\end{equation}
where we used $\int \dX p(\vecX) \ln \Gamma(\vecX)=0$ upon enforcement of the constraint. 

The derivation above shows that $p(\vecX)$ is a critical point for the entropy. We now show that $p(\vecX)$ is indeed a maximum. Assume that the true distribution is $p_0(\vecX)$, which is  normalized and satisfies Eq.~\eqref{eq:MegaCF}. Using the inequality $\ln x-1+1/x\geq0$ for any $x>0$, we first notice that $\int \dX p_0(\vecX)\ln[p_0(\vecX)/p(\vecX)]\geq\int \dX p_0(\vecX)[1-p(\vecX)/p_0(\vecX)]=0$. Thus, the thermodynamic excess entropy is bound according to
\begin{eqnarray}
    S-S^\mathrm{id}&=&-\int \dX p_0(\vecX) \ln p_0(\vecX)\leq-\int \dX p_0(\vecX) \ln p(\vecX)\nonumber\\
    &=&\int \dX p_0(\vecX) \ln\z-\int \dX p_0(\vecX) \ln \Gamma(\vecX)=\ln Z=H_\mathrm{ex}.
\end{eqnarray}
where we have used $\int \dX p_0(\vecX) \ln \Gamma(\vecX)=0$ since $p_0(\vecX)$ gives rise to the same constrained CF. This proof holds for any additional constraints (including those which we introduce in the SM~\cite{SM}).

Since $\vecB\bphi$ is Hermitian by construction, and the Lagrangian is real, only the Hermitian part of $\blam$ survives. In addition, transposing $\blam$ amounts to relabeling the constraints. Henceforth, we will drop the $\dagger$. 

Thus we have found the relation between the bound on the excess entropy $H_\mathrm{ex}$ and the effective partition function $\z$, and between $\z$ and the Lagrange multiplier $\blam$ and the constraint $\bphi$. The procedure continues as follows. (a) Perform the integral of Eq.~\eqref{eq:partition} explicitly, and find $\z$ as a functional of $\bphi$ and $\blam$. (b) Obtain $\blam$ as a function of $\bphi$ using Eq.~\eqref{eq:enforceZ} and, consequently, find $\z$ in terms of $\bphi$ alone. (c) Obtain $H$ (Eq.~\eqref{eq:entpart}) as a functional of $\bphi$. A similar calculation was carried out in Ref.~\cite{ArielPRE2020}. There, $\bphi$ was limited to the number density field, as the only DOF was the particles' locations, and the CF was the structure factor. Here, $\bphi$ is the field of a general DOF $\vecj$. In the SM~\cite{SM} we also include a steady inhomogeneous field and a varying number of particles.

Similar to Ref.~\cite{ArielPRE2020} we transform to integration over fields, $\vecJ=\sum_n\vecg(\vecx_n)$, which we introduce through a functional delta,
\begin{equation}
    \delta\left[\vecJ-\sum_n\vecg(\vecx_n)\right]=(2\pi)^{-\mt}\int D\bpsi\exp\left[ i\bpsi^\dagger\left(\vecJ-\sum_n\vecg(\vecx_n)\right) \right],
\end{equation}
where $\bpsi$ (a $\mt$-component column vector) is the conjugate field. These fields are no longer functions of individual particles' positions, but rather functions of the modes. For later convenience, we add and subtract the uniform mode, which by construction satisfies $\vecJ_0=\sum_{n}\vecg_0(\vecx_n)=N$ for any $\vecx_n$. Some arbitrary value for the corresponding $\bpsi_0$ may be assumed. The partition function becomes
\begin{equation}
    \z=(2\pi)^{-\mt}e^{\tr\,\blam\vecB\bphi}\int \dX D\vecJ D\bpsi \exp\left[i\tilde{\bpsi}^\dagger\left(\tilde{\vecJ}-\sum_n\tilde{\vecg}(\vecx_n)\right)-\vecJ^\dagger\blam\vecJ \right],
\end{equation}
where the tilde indicates that $\tilde{\vecg}(\vecx)$ is the complete transform (\ie including the uniform mode which we have been excluding thus far). Therefore, $\tilde{\vecg}(\vecx)$, $\tilde{\vecJ}$, and $\tilde{\bpsi}$ have $\mt+1$ components. The ``tilde-less'' $\vecg$, $\vecJ$, and $\bpsi$ still exclude the uniform mode, and as before are of size $\mt$. The integral over the configurations can be rewritten as
\begin{align}
    \int \dX \exp\left(-i\tilde{\bpsi}^\dagger\sum_{n=1}^N\tilde{\vecg}(\vecx_n)\right)=\left[\int \dx e^{-i\tilde{\bpsi}^\dagger\tilde{\vecg}(\vecx)}\right]^N=\left[1+\int \dx\left( e^{-i\tilde{\bpsi}^\dagger\tilde{\vecg}(\vecx)}-1\right)\right]^N \nonumber\\ \xrightarrow[]{N\to\infty,\prob(\vecx)\sim\upsilon^{-1}\to0,N\dx\sim1;\tilde{\bpsi}\to0}\exp\left[N\int \dx \left(e^{-i\tilde{\bpsi}^\dagger\tilde{\vecg}(\vecx)}-1\right)\right].
\end{align}
We are left with the field integrals,
\begin{equation}
    \z=(2\pi)^{-\mt}e^{\tr\,\blam\vecB\bphi}\int D\vecJ D\bpsi \exp\left[N\int \dx \left(e^{-i\tilde{\bpsi}^\dagger\tilde{\vecg}(\vecx)}-1\right)+i\tilde{\bpsi}^\dagger\tilde{\vecJ}-\vecJ^\dagger\blam\vecJ \right].
\end{equation}

One course of action is to first perform the exact Gaussian integral over $\vecJ$ and remain with a non-Gaussian integral over $\bpsi$. The next step in Ref.~\cite{ArielPRE2020} was to take the Gaussian approximation for this integral, thus including the leading-order fluctuations. An alternative course of action (involving the same level of approximation) is to first perform a saddle-point approximation for $\bpsi$, and then take a Gaussian approximation for $\vecJ$. For the sake of this Appendix, both methods are equally convenient, but in the SM~\cite{SM} (including the constraint on a steady profile $\vecG$), the latter turns out to be much easier. The former method could be found in Ref.~\cite{ArielPRE2020}, and the latter is shown below. These approximations\,---\,saddle point in $\bpsi$ and Gaussian in $\vecJ$\,---\,are the only ones employed. 

We rewrite the partition function as
\begin{subequations}
\begin{eqnarray}
    \z&=&(2\pi)^{-\mt}e^{\tr\,\blam\vecB\bphi}\int D\vecJ e^{-\vecJ^\dagger\blam\vecJ}\int D\bpsi e^{-F[\tilde{\bpsi};\tilde{\vecJ}]},
    \label{eq:Zsp}\\
    F[\tilde{\bpsi};\tilde{\vecJ}]&=&-N\int \dx \left(e^{-i\tilde{\bpsi}^\dagger\tilde{\vecg}(\vecx)}-1\right)-i\tilde{\bpsi}^\dagger\tilde{\vecJ}.\label{eq:frenergy}
\end{eqnarray}
\end{subequations}
In the saddle point approximation, we seek $\tilde{\bpsi}=\tilde{\bpsi}_\mathrm{sp}(\tilde{\vecJ})$ that minimizes the effective free energy, $F[\tilde{\bpsi}_\mathrm{sp}(\tilde{\vecJ});\tilde{\vecJ}]$, for each $\tilde{\vecJ}$, and then take leading order corrections. The equation for $\tilde{\bpsi}_\mathrm{sp}$ is
\begin{equation}
    i\frac{\partial F}{\partial\tilde{\bpsi}^\dagger}=\bbo=\tilde{\vecJ}-N\int \dx\tilde{\vecg}(\vecx)e^{-i\tilde{\bpsi}^\dagger_\mathrm{sp}\tilde{\vecg}(\vecx)}.\label{eq:psiSP}
\end{equation}

Since the transform $\tilde{\vecg}(\vecx)$ is a complete basis, there exists the dual object $\tilde{\vecb}(\vecx)$ such that $\tilde{\vecb}^{\dagger}(\vecx)\tilde{\vecg}(\vecx')=\delta(\vecx-\vecx')/\prob(\vecx)$ (and $\int \dx \tilde{\vecg}(\vecx)\tilde{\vecb}^\dagger(\vecx)=\tilde{\mathcal{I}}$, being the $(\mt+1)\times(\mt+1)$ unit matrix), where again $\prob(\vecx)$ serves as the weight function. As an example, consider the simple particle-density field, where $\tilde{\vecg}_\vecq(\vecr)=e^{-i\vecq\cdot\vecr}$, such that $\tilde{\vecb}_\vecq(\vecr)=e^{-i\vecq\cdot\vecr}$ and $\prob(\vecr)=1/V$, and, indeed, $\sum_\vecq \tilde{\vecb}^*_\vecq(\vecr)\tilde{\vecg}_\vecq(\vecr)=\delta(\vecr-\vecr')/\prob(\vecr)$ (and $\int d\vecr \prob(\vecr) \tilde{\vecg}_\vecq(\vecr)\tilde{\vecb}^*_{\vecq'}(\vecr)=\delta_{\vecq\vecq'}$). With the help of $\tilde{\vecb}(\vecx)$, we can solve Eq.~\eqref{eq:psiSP} as
\begin{equation}
    \tilde{\bpsi}^\dagger_\mathrm{sp}(\tilde{\vecJ})=i\int \dx\ln[N^{-1}\tilde{\vecb}^\dagger(\vecx)\tilde{\vecJ}]\tilde{\vecb}^\dagger(\vecx).
\end{equation}

We now expand the free energy around the saddle point, $\tilde{\bpsi}_\mathrm{sp}$,
\begin{subequations}
\begin{eqnarray}
    F[\tilde{\bpsi}_\mathrm{sp}(\tilde{\vecJ});\tilde{\vecJ}]&=&N\int \dx \left[\frac{\tilde{\vecb}^\dagger(\vecx)\tilde{\vecJ}}{N}\ln\left(\frac{\tilde{\vecb}^\dagger(\vecx)\tilde{\vecJ}}{N}\right)+1-\frac{\tilde{\vecb}^\dagger(\vecx)\tilde{\vecJ}}{N}\right] \simeq\frac{1}{2}\vecJ^\dagger\vecB^{-1}\vecJ,\label{eq:SP0th}\\
    \frac{\delta^2F}{\delta\bpsi^\dagger\delta\bpsi}&=&\int \dx \tilde{\vecg}(\vecx)\tilde{\vecg}^\dagger(\vecx)\tilde{\vecb}^\dagger(\vecx)\tilde{\vecJ}
    \simeq\vecB.
\end{eqnarray}
\end{subequations}
We have kept terms to quadratic order in $\bpsi$ and $\vecJ$, $\vecJ_0$ has naturally disappeared, and fluctuations in $\bpsi_0$ are irrelevant. The approximate free energy reads
\begin{equation}
    F[\bpsi;\vecJ]\simeq\frac{1}{2}\vecJ^\dagger\vecB^{-1}\vecJ+\frac{1}{2}\bpsi^\dagger\vecB\bpsi.
    \label{eq:Fsp}
\end{equation}
Substituting Eq.~(\ref{eq:Fsp}) in Eq.~(\ref{eq:Zsp}), we obtain the partition function
\begin{eqnarray}
    \z&\simeq&(2\pi)^{-\mt}e^{\tr\,\blam\vecB\bphi}\int D\vecJ e^{-\frac{1}{2}\vecJ^\dagger\left(2\blam+\vecB^{-1}\right)\vecJ}\int D\bpsi e^{-\frac{1}{2}\bpsi^\dagger\vecB\bpsi}\nonumber\\
    &=&\exp\left(\tr\left[\blam\vecB\bphi-\frac{1}{2}\ln\left(2\blam\vecB+\mathcal{I}\right)\right]\right),\label{eq:finalparthom}
\end{eqnarray}
where we have used the matrix property, $\det[\exp(\cdot)]=\exp[\tr(\cdot)]$. 

Thus we have found $\z$ as a functional of $\blam$ and $\bphi$. From Eq.~\eqref{eq:enforceZ}, we obtain the effective pair-potential,
\begin{equation}
    \blam\simeq\frac{1}{2}\left(\bphi^{-1}-\mathcal{I}\right)\vecB^{-1}.\label{eq:pairphom}
\end{equation}
Substitution of Eq.~\eqref{eq:pairphom} in Eq.~\eqref{eq:finalparthom}, and the result in Eq.~\eqref{eq:entpart} gives,
\begin{equation}
    h_{\rm ex}[\bphi] \simeq \frac{1}{2N} \tr\left[\ln\bphi+\mathcal{I}-\bphi\right],
\end{equation}
which is the desired bound, Eq.~\eqref{eq:hent} (with $\vecG=\bbo$ and $\Nmean=N$). In the SM~\cite{SM}, we derive the bound for a nonzero $\vecG$, and for the grand-canonical ensemble.\vphantom{\cite{Book:FrenkelSmit}}

\newpage
\renewcommand{\theequation}{S\arabic{equation}}
\setcounter{equation}{0}
\renewcommand{\thefigure}{S\arabic{figure}}
\setcounter{figure}{0}
\renewcommand{\thetable}{S\arabic{table}}
\setcounter{table}{0}
\renewcommand{\thesection}{\Roman{section}}
\setcounter{section}{0}

\section*{Resolving entropy contributions in nonequilibrium transitions: Supplementary Material}

Below we provide further information on which we did not elaborate in the main text. Further information regarding the analysis of the bidisperse mixtures experiment~\cite{ricouvier2017optimizing} (Sec.~III of the main text) is given in Sec.~\ref{appendix_joshua}. There, we explain how we constructed the structure factor, spectral density $\chi(\vecq)$, and their mixed correlation function (CF). We also attach snapshots and spectral densities for more number fractions (\textit{NF}s) and size ratios (\textit{SR}s). In Sec.~\ref{appendix:enderiv} we complete the derivation of Eq.~(A3) of the main text; in Sec.~\ref{appendix:profile}
we additionally impose inhomogeneity, and in Sec.~\ref{appendix:grand} we let the number of particles vary. We follow the same notation adopted in the main text. Finally, in Sec.~\ref{appendix:example} we demonstrate through a toy model the failure of various computational methods for entropy estimation in the case of a variable number of particles.

\section{Constructing the correlation functions}
\label{appendix_CFb}

\subsection{Correlation functions in the Vicsek model}
\label{appendix_CFvicsek}
As explained in Sec. III of the main text, we construct three CFs: the structure factor, the orientational CF, and a mixed CF. Below we describe how to compute them from simulation data and provide accurate expressions for the excess entropies.

\begin{itemize}
\item 
The structure factor, $S(\vecq)$ is formally defined by taking $\vecf(\theta)=1$ in Eqs.~(2) of the main text. 
For a real scalar field, the normalization factor $\vecA$, given by Eq.~(3b) of the main text, is just the 2nd moment of $\vecf(\theta)$, hence $\vecA=1$. 

For a periodic domain $[0,L]^2$, the Fourier modes are discrete,
$\vecq=2\pi/L(n_x,n_y)$, $n_x,n_y\in \mathbb{Z}$. We apply a cutoff at a maximal mode $|n_x|,|n_y| \le \Omega=50$, chosen such that for the higher modes all CFs become equal to 1 up to statistical noise. Next, the discrete field of a sample $s$, $\vecJ^s_0(\vecq)=\sum_{n=1}^Ne^{-i\vecq\cdot \vecr_n^s}$ can be computed (note an $O(N)$ computational cost). The ensemble averaged structure factor, for each $\vecq$, is then given by
\begin{equation}
    S(\vecq)=\frac{1}{N}\langle|\vecJ^s_0(\vecq)|^2\rangle_s 
    = \frac{1}{N K} \sum_{s=1}^K | \vecJ^s_0(\vecq)|^2,
\end{equation}
where $K$ is the number of samples (here $10^4$). 
Finally, the excess entropy, Eq.~(4) of the main text, is
\begin{equation}
  h_\mathrm{ex} [S] 
  = \frac{1}{N} \sum_{n_x=1}^\Omega \sum_{n_y=-\Omega}^{\Omega} [\ln S(\vecq)+1-S(\vecq)].
\end{equation}
Note the summation range of $n_x$ and $n_y$,
where we used the fact that $S(\vecq)=S(-\vecq)$.
\item
Orientational CF, $\vecD(\vecq)$\,---\,defined with $\vecf(\theta)=(\cos\theta,\sin\theta)^T$. Since the orientations are annealed, there is no need to compute their marginal distributions. The normalization matrix $\vecA$ is the 2nd moment matrix for $\vecf$, 
\begin{equation}
  \vecA = \langle \vecf \vecf^\dag \rangle =
  \left\langle \left( \begin{array}{c} \cos\theta \\ \sin\theta \end{array} \right) 
  \begin{array}{c} \left( \cos\theta , \sin\theta\right)  \\ \, \end{array}
  \right\rangle = \left( \begin{tabular}{cc} 1/2 & 0 \\ 0 & 1/2 \end{tabular} \right).
\end{equation}
The Fourier mode discretization is the same as above. 
Next, for each sample $s$, we compute the discrete fields,
\begin{equation}
\vecJ_1^s(\vecq)=\sum_{n=1}^Ne^{-i\vecq\cdot \vecr_n^s}\cos(\theta_n^s) ~,~~~~  \vecJ_2^s(\vecq)=\sum_{n=1}^Ne^{-i\vecq\cdot \vecr_n^s}\sin(\theta_n^s). 
\end{equation}
The CF $\vecD$ is a $2 \times 2$ matrix with components 
\begin{equation}
\vecD_{\alpha\beta}=\frac{2}{N} \langle\vecJ_\alpha^s(\vecq)(\vecJ_\beta^s)^*(\vecq)\rangle_s ,
\end{equation}
where $\alpha,\beta=1,2$. 
Substituting into Eq.~(4) of the main text, and using $\tr \ln [\cdot] = \ln \det [\cdot]$, the excess entropy associated with $\vecD$ is
\begin{equation}
h_\mathrm{ex} [\vecD]=\frac{1}{N} \sum_{n_x=1}^\Omega \sum_{n_y=-\Omega}^{\Omega} [\ln(\vecD_{11}(\vecq)\vecD_{22}(\vecq)-|\vecD_{12}(\vecq)|^2)+2-\vecD_{11}(\vecq)-\vecD_{22}(\vecq)].
\end{equation}
\item
Mixed CF, $\vecM(\vecq)$\,---\,we seek to impose simultaneously both $S\propto\langle|\vecJ^s_0|^2\rangle_s$ and $\vecD_{\alpha\beta}\propto\langle|\vecJ_\alpha^s(\vecJ_\beta^s)^*|^2\rangle_s$. Therefore, we define the mixed CF with $\vecf(\theta)=(1,\cos\theta,\sin\theta)$. Similar to the above, the normalization matrix is $\vecA=\mathrm{diag}(1,1/2,1/2)$. The CF is a $3\times3$ matrix with components, $\vecM_{ij}(\vecq)=(\vecA_{ii}N)^{-1}\langle\vecJ_i^s(\vecq)(\vecJ_j^s)^*(\vecq)\rangle_s$, $i,j=0,1,2$. Note that in addition to the previous two CFs (located as diagonal blocks in $\vecM$), $\vecM_{01}=\vecM_{10}^*$ and $\vecM_{02}=\vecM_{20}^*$ are the position-orientation cross-correlations. Similar to the above, the excess entropy is, \begin{equation}
h_\mathrm{ex} [\vecM] =\frac{1}{N} \sum_{n_x=1}^\Omega \sum_{n_y=-\Omega}^{\Omega} [\ln(\det\vecM(\vecq))+3-\tr\vecM(\vecq)].
\end{equation}
\end{itemize}

\subsection{Correlation functions in the bidisperse mixture}
\label{appendix_CFjoshua}

Unlike the results from most simulated data, which are typically organized in a list containing the particles' degrees of freedom (DOFs), the experimental data are snapshots. For this reason, it is more convenient and efficient to process the snapshots directly. For each snapshot, which is one statistical sample $s$, we do the following: 
\begin{enumerate}
    \item Each snapshot is a matrix of pixel intensities. Every disk-shaped collection of pixels corresponding to a single droplet is replaced with a single pixel at the disc's center of mass, carrying the value of the disc's area (the rest are zeroed). We denote the resulting matrix $M_s$.
    \item We proceed to find the DOF fields (Eq.~(2a) of the main text). For the density field (with $\vecf(a)=1$), denoted $\vecJ_{0,s}(\vecq)$, we replace all non zero entries in $M_s$ (which are disk areas) with the value 1. The matrix is Fourier-transformed (using 2D fast Fourier transform). For the areas field (with $\vecf(a)=a$), denoted $\vecJ_{1,s}(\vecq)$, we Fourier-transform $M_s$.
    \item\label{step:norm} Normalization. The number of droplets slightly varies between snapshots. Therefore, for each sample, we compute the number of particles, $N_s$ (the number of non-zero entries in $M_s$). We will need the average area $\overline{a}_s$ (the average of all non-zero entries in $M_s$) and similarly the second moment $\baa_s$. $N_s$ and $N_s\baa_s$ will be used to normalize the structure factor and the spectral density (areas CF), respectively, and the following
    matrix will normalize the mixed CF:
    \begin{equation}
        \vecA_{\vecM_s} = \left( \begin{tabular}{c c}
             1 &  $\overline{a}_s$ \\
             $\overline{a}_s$ & $\baa_s$
        \end{tabular} \right) .
    \end{equation}
    \item\label{step:CF} For each snapshot, we compute three normalized CFs (Eqs.~(3) of the main text): (a) The structure factor\,---\,$S_s(\vecq)=|\vecJ_{0,s}(\vecq)|^2/N_s$. (b) The spectral density, $\chi^s(\vecq)=|\vecJ_{1,s}(\vecq)|^2/(N_s\baa_s)$. (c) A mixed CF, $\vecM_s(\vecq)$, imposing the two CFs simultaneously along with their cross-correlations. For every $\vecq$, it is a $2\times2$ matrix:
    \begin{equation}
        \vecM_s(\vecq) = \frac{1}{N_s} \vecA_{\vecM_s}^{-1} 
        \left( \begin{tabular}{c}
             $\vecJ_{0,s}(\vecq)$  \\ $\vecJ_{1,s}(\vecq)$
        \end{tabular} \right)
        \begin{tabular}{c c}
            $\left( \vecJ_{0,s}^*(\vecq)\quad \vecJ_{1,s}^*(\vecq)  \right)$ 
        \end{tabular}
    \end{equation}
    \item Assume radial symmetry. Partition the range $[0,q_{\rm max}]$ into equal bins. For each bin $(q_1,q_2]$, average each correlation function for all $\vecq$ with a norm in that range. Note that the $\vecq=\veco$ mode is not included.
    \item Substituting into the expression for the excess entropy (Eq.~(4) of the main text), sum over $q>0$ with an additional multiplicative term $2 \pi q$ for the Jacobian (because the summation is over the radial direction).
\end{enumerate}

\section{Derivation of generalizations}
\label{appendix:enderiv}

\subsection{Entropy bound with a given steady profile}
\label{appendix:profile}

Building upon the derivation of Appendix~B of the main text, we include another constraint in addition to the CF and probability normalization, namely, an inhomogeneous stationary profile,
\begin{equation}
    \vecG=\frac{1}{\sqrt{N}}\int \dX p(\vecX)\sum_{n=1}^N\vecg(\vecx_n).\label{eq:MegaG}
\end{equation}
This would lower the entropy bound further for inhomogeneous systems.\footnote{The imposed profile $\vecG$ is not equivalent to the single-particle distribution $\Pr(\vecx)$. The latter is defined as the distribution in the absence of any external perturbation, whereas the actual single-particle distribution will be affected by interactions and external fields (leading to the profile $\vecG$).} 

With that additional constraint, we redefine the Lagrangian as
\begin{eqnarray}
  {\mathcal L} &=&-\int \dX p(\vecX) \ln p(\vecX)- \eta \left( \int \dX p(\vecX) - 1 \right) \nonumber\\
   &&- \tr\left[ \blam \int \dX p(\vecX)\left(\sum_{n,m}\vecg(\vecx_n)\vecg^\dagger(\vecx_m) - \vecB \bphi \right) \right]\nonumber \\ &&-\bmu^\dagger\int \dX p(\vecX)\left(\sum_{n}\vecg(\vecx_n) - \sqrt{N} \vecG \right),
\end{eqnarray}
where $\bmu$ (a $\mt$-component column vector) is the additional Lagrange multiplier for the profile. We have, in addition to Eqs.~(B7) of the main text, 
\begin{equation}
   \frac{\partial {\mathcal L} }{\partial \bmu^\dagger} = \bbo.
\end{equation}

The probability distribution, $p(\vecX)=\z^{-1}\Gamma(\vecX)$, is modified to
\begin{equation}
   \Gamma(\vecX)=\exp\left(-\tr\left[ \blam \left(\sum_{n,m}\vecg(\vecx_n)\vecg^\dagger(\vecx_m) - \vecB \bphi \right) \right]-\bmu^\dagger\left(\sum_{n}\vecg(\vecx_n) - \sqrt{N} \vecG \right) \right),
\end{equation}
which is again a Boltzmann-like factor, with the yet unknown effective external field $\bmu$. The effective partition function is still $\z=\int\dX\Gamma(\vecX)$. Similar to Eq.~(B10) of the main text,
\begin{equation}
    \frac{\partial\ln\z}{\partial\bmu^\dagger}=\bbo\label{eq:enforcemu}
\end{equation}
is equivalent to enforcing the profile constraint. The entropy bound is still related to the partition function through $H=\ln\z$.

The calculation procedure is identical to the one presented in Appendix~B of the main text, with the addition of finding $\bmu$. After the transition to integrals over fields, the partition function becomes
\begin{subequations}
\begin{eqnarray}
  \z&=&\int \dX e^{-\tr\left[ \blam \left(\sum_{n,m}\vecg(\vecx_n)\vecg^\dagger(\vecx_m) - \vecB \bphi \right) \right] -\bmu^\dagger\left(\sum_{n}\vecg(\vecx_n) - \sqrt{N} \vecG \right)} \\
  &=&(2\pi)^{-\mt}e^{\tr\,\blam\vecB\bphi+\sqrt{N}\bmu^\dagger\vecG}\times \nonumber \\ 
  &&\times\int D\vecJ D\bpsi \exp\left[N\int \dx \left(e^{-i\tilde{\bpsi}^\dagger_0\tilde{\vecg}(\vecx)}-1\right)+i\tilde{\bpsi}^\dagger\tilde{\vecJ}-\vecJ^\dagger\blam\vecJ-\bmu^\dagger\vecJ \right].
\end{eqnarray}
\end{subequations}
We see that the effective free energy, defined for the sake of a saddle point approximation, remains the same as in Appendix~B of the main text (Eq.~(B16b) of the main text). Therefore, following the same steps, we obtain the approximate partition function
\begin{eqnarray}
    \z&\simeq&(2\pi)^{-\mt}e^{\tr\,\blam\vecB\bphi+\sqrt{N}\bmu^\dagger\vecG}\int D\vecJ e^{-\frac{1}{2}\vecJ^\dagger\left(2\blam+\vecB^{-1}\right)\vecJ-\bmu^\dagger\vecJ}\int D\bpsi e^{-\frac{1}{2}\bpsi^\dagger\vecB\bpsi}\nonumber\\
    &=&\exp\left(\tr\left[\blam\vecB\bphi-\frac{1}{2}\ln\left(2\blam\vecB+\mathcal{I}\right)\right]+\sqrt{N}\bmu^\dagger\vecG+\frac{1}{2} \bmu^\dagger\left(2\blam+\vecB^{-1}\right)^{-1}\bmu\right).
\end{eqnarray}

We now impose the profile constraint (Eq.~\ref{eq:enforcemu}), obtaining
\begin{subequations}
\begin{equation}
    \bmu=-\sqrt{N}\left(2\blam+\vecB^{-1}\right)\vecG,
\end{equation}
and the CF constraint  (Eq.~(B10) of the main text), leading to
\begin{equation}
    \blam=\frac{1}{2}\left[\left(\bphi-N\vecB^{-1}\vecG\vecG^\dagger\right)^{-1}-\mathcal{I}\right]\vecB^{-1}.
\end{equation}
\end{subequations}
The modified expression for the entropy's upper bound becomes
\begin{equation}
    h_{\rm ex}[\bphi,\vecG]=\frac{1}{2N}\tr\left[\ln\left(\bphi-\vecA^{-1}\vecG\vecG^\dagger\right)+\mathcal{I}-\bphi\right],
\label{eq:hexG}
\end{equation}
which coincides with Eq.~(A3) of the main text (with $\langle N\rangle = N$).

\subsection{Entropy bound with varying number of particles}
\label{appendix:grand}

In Appendix~B of the main text and Sec.~\ref{appendix:profile} here we computed the entropy bound resulting from measured CF and steady profile, for a given number of particles $N$. In this Section we generalize the results for a grand-canonical ensemble of identical particles. Assume a system in which the probability to have $N$ particles is $P(N)$. The joint density $P(\vecX,N)$ is defined as,
\begin{equation}
	P(\vecX,N) = P(N) \prod_{n=1}^N \prob(\vecx_n) p( \vecX | N),
\label{eq:PNX}
\end{equation}
where $p( \vecX | N)$ is $p(\vecX)$ of Appendix~B of the main text and Sec.~\ref{appendix:profile} here for a fixed $N$. Accordingly, the Shannon entropy of the entire ensemble spits into three contributions,
\begin{eqnarray}
	H &=& - \sum_{N=0}^\infty \int d\vecX P(N,\vecX) \ln P(N,\vecX) \nonumber\\
	&=& H_N + \Nmean h_1 + \sum_{N=0}^\infty P(N)Nh_\mathrm{ex}(N) ,
\label{eq:HNX}
\end{eqnarray}
where $H_N=-\sum_{N=0}^\infty P(N) \ln P(N)$ is the entropy associated with the system's occupancy $N$, $h_1=$\linebreak$-\int d\vecx \prob(\vecx)\ln\prob(\vecx)$ is the single-particle configurational entropy of Eq.~(B4) of the main text, and $h_\mathrm{ex}(N)$ is the excess entropy of configurations $\vecX$ at given $N$, Eq.~\eqref{eq:hexG}. Assuming a Gaussian distribution for $N$ with average $\Nmean$ and variance $\sigma^2 \Nmean$, $H_N =\frac12 \ln (2 \pi e \sigma^2 \Nmean)$. Therefore, in the absence of correlations, the entropy is
\begin{equation}
    H^\mathrm{id}=H_N + \Nmean h_1=\frac12 \ln (2 \pi e \sigma^2 \Nmean)-\Nmean \int d\vecx \prob(\vecx)\ln\prob(\vecx).
\end{equation}

Applying Eq.~\eqref{eq:HNX} directly to calculate the grand-canonical entropy $H$ is impractical. Typical statistical entropy estimators require an order of $2^{dN}$ independent samples \cite{Kozachenko1987,Beirlant1997review,Darbellay1999,Paninski2003,Stowell2009,Lord2018,Ariel2020}. Even when multiple realizations are available, the number of independent samples {\em for each $N$} will typically be much lower. Moreover, combining samples with different dimensions is not straightforward. Typical statistical entropy estimators assume that the samples are vectors with a fixed dimension~\cite{Kozachenko1987,Beirlant1997review,Darbellay1999,Paninski2003,Stowell2009,Lord2018,Ariel2020}. In other methods which do not explicitly assume that samples are independent~\cite{AvineryPRL2019,MartinianiPRX2019}, realizations with a different number of particles can be artificially concatenated or trimmed to produce a single long sequence whose information content can be estimated. However, mixing samples of different sizes can ruin the frequency of recurrent patterns, reducing the accuracy of such methods. In Sec.~\ref{appendix:example} we give a numerical example of such failures.

In the following, we assume that the average and variance of the number of particles, $\Nmean$ and $\sigma^2 \Nmean$, the two-point CF $\bphi$, and the steady profile $\vecG$, are given. In preceding sections, we showed that the excess entropy per particle $h_\mathrm{ex}(N) = H_\mathrm{ex}(N)/N$ can be approximately bounded by
\begin{equation}
	h_\mathrm{ex}(N)=\frac{1}{2N}\tr\left[\ln\left(\bphi(N)-\vecA^{-1}\vecG(N)\vecG(N)^\dagger\right)+\mathcal{I}-\bphi(N)\right],
\end{equation}
where $\bphi(N)$ and $\vecG(N)$ are the CF and profile for a given $N$. The CF and profile as defined in Eq.~(B3) of the main text and Eq.~\eqref{eq:MegaG} are intensive properties; thus they are not expected to vary significantly with $N$ for sufficiently large $N$. Therefore, to leading order in $\Nmean$, $\bphi(N)$ and $\vecG(N)$ may be replaced with their average values $\bphi = \sum_N P(N) \bphi(N)$ and $\vecG=\sum_N P(N) \vecG(N)$, respectively. Recalling that the trace includes a multiplicative volume dependence through the $\vecq$ discretization, the intensive $h_\mathrm{ex}(N)$ also does not depend strongly on $N$, and the overall excess entropy becomes, to leading order,
\begin{equation}
	H_\mathrm{ex} = H-H^\mathrm{id} = H_N - \frac12 \ln (2 \pi e \sigma^2 \Nmean)   + \frac{1}{2}\tr\left[\ln\left(\bphi-\vecA^{-1}\vecG\vecG^\dagger\right)+\mathcal{I}-\bphi\right].
\end{equation}
We recall that the entropy-maximizing distribution,  given its variance only, is Gaussian. Hence, the first two terms cancel and we obtain Eq.~(A3) of the main text,
\begin{equation}
	h_\mathrm{ex} =  \frac{1}{2\Nmean}\tr\left[\ln\left(\bphi-\vecA^{-1}\vecG\vecG^\dagger\right)+\mathcal{I}-\bphi\right].
\label{eq:hNX}
\end{equation}
Unsurprisingly, the excess entropy of the grand-canonical ensemble is the same as that of the canonical one. Yet, this simple understanding implies that Eq.~(A3) of the main text can be used to calculate the excess entropy of systems in which the number of particles is not constant.

\section{Estimation failures with variable dimensions}
\label{appendix:example}

To test the applicability of various entropy estimation methods for grand-canonical ensembles, we apply them to the equilibrium distribution of the following simple 1D model.
The model consists of $N$ particles, connected by linear springs with fixed spring constant $K$ and relaxated length $L$.
In addition, each particle is connected to the same point (the origin) by a linear spring with spring constant $1$ and relaxed length $0$. The Hamiltonian is
\begin{equation}
  H = \frac{K}{2} \sum_{i \neq j} (x_i-x_j-L)^2 + \frac{1}{2} \sum_i x_i^2 ,
\end{equation}
where $x_i \in \mathbb{R}$ is the position of the $i$th particle.
The equilibrium distribution is sampled using standard Monte-Carlo sampling \cite{Book:FrenkelSmit}. We take $100N$ MC relaxation steps between samples, attempting to move the position of one randomly chosen particle by $\mathcal{N}(0,\Delta^2)$, $\Delta=0.5 L$.
Canonical ensembles are simulated with fixed inverse temperature $\beta$ and fixed number of particles $N$.
Grand-canonical ensembles are simulated with fixed inverse temperature $\beta$ and fixed activity $z=e^{\beta \mu}$, where $\mu$ is the chemical potential.
Writing the system as a Langevin dynamics, we obtain a linear stochastic differential equation with a constant diffusion term. As a result, the invariant distribution is Gaussian.

Results for the total entropy per particle of a canonical ensemble with $\beta=0.5$, $K=1$, and $L=1$, using $10^5$ samples, are given in Table~\ref{tbl:Cresults100k}. 
The number of particles is quite small (up to $N=200$), as direct sampling methods cannot be applied for high dimensions. The results are not sensitive to the number of samples.
The column labeled `Gaussian' shows the entropy (per particle) of a Gaussian multivariate distribution with covariance matrix obtained numerically from the simulation.
The column labeled `copula' shows results using the recursive copula splitting method introduced in Ref.~\cite{Ariel2020}.
The column labeled `compression' shows results using the compression-based method introduced in Refs.~\cite{AvineryPRL2019,MartinianiPRX2019}. Implementation details follow Ref.~\cite{ArielPRE2020}.
The copula method is quite accurate, with accuracy of up to $20\%$ compared to the Gaussian estimate.
The estimate of the compression-based method is of the correct scale, but is significantly less accurate, with errors up to $333\%$.

\begin{table}[ht]
\centering{}%
\begin{tabular}{|c|c|c|c|}
\hline
$N$  & Gaussian (exact) & copula & compression  \\
\hline
$2$ & 1. & 0.99 & 0.91 \\
$5$ & 0.98 & 0.86 & 0.4 \\
$10$ & 0.94 & 0.9 & 0.1 \\
$20$ & 0.91 & 0.79 & -0.06 \\
$50$ & 0.89 & 0.75 & 0.12 \\
$100$ & 0.88 & 0.73 & 0.59 \\
$200$ & 0.87 & 0.72 & 1.56 \\
\hline
\end{tabular}
\caption{Canonical simulations with $10^5$ samples.}
\label{tbl:Cresults100k}
\end{table}

Table~\ref{tbl:GCresults100k} shows the results for the grand-canonical ensemble.
In order to apply the copula method, the data need to be transformed into independent vectors of fixed dimension. We tried three strategies.
The first, labeled `split $N$', splits the data into the different values for $N$ and estimates the entropy for each case independently.
Then, the weighted average is taken.
The second strategy, labeled `concat', concatenates all samples into a long vector and reshapes it into a $M \times \langle N \rangle$ matrix, where $M$ is the number of samples.
The third strategy, labeled `truncate', randomly removes samples so they all have the same dimension (here, the median $N$).
Samples with smaller $N$ are discarded. (Applying the cutoff at a different precentile does not change the results qualitatively.)
Most of the error (about $200\%$) comes from the concatenation/truncation procedure.  
The error of the copula method (compared to the Gaussian estimate) is significantly larger compared to the canonical ensemble, with errors up to $7\%$.
The compression-based method is off by two-three orders of magnitude.

\begin{table}[h]
\centering{}%
\begin{tabular}{|c|c||c|c|c||c|c|c||c|c|c|}
\hline
 & & split $N$ & & & concat & & & truncate & & \\
$z$ & $N \pm \Delta N$  & Gaussian & copula & compress & Gaussian & copula & compress & Gaussian & copula & compress  \\
\hline
$0$ & $3.4 \pm 2.5 $ & 0.75 & 0.13 & -1.5  & -0.8 & -2.1 & -2.5 & -0.7 & -2.1 & -2.6 \\
$1$ & $9.4 \pm 2.8 $ & -0.08 & -1.5 & -25  & -0.9 & -2.4 & -23 & -0.9 & -2.4 & -23 \\
$5$ & $25 \pm 3.5 $ & -0.9 & -2.8 & -96  & -1.4 & -3.1 & -95 & -1.4 & -3.1 & -90 \\
$20$ & $74 \pm 8.4 $ & -1.9 & -4.1 & -391  & -2.3 & -4.4 & -387 & -2.3 & -4.4 & -374 \\
$50$ & $180 \pm 21 $ & -2.7 & -5.0 & -1106  & -3.1 & -5.5 & -1128 & -3.1 & -5.5 & -1104 \\
\hline
\end{tabular}
\caption{Grand-canonical simulations with $10^5$ samples.}
\label{tbl:GCresults100k}
\end{table}

Overall, we see that standard entropy estimation methods cannot be blindly applied to grand-canonical ensembles.
This is not surprising taking into account that such methods assume a fixed-dimensional probability space.


\begin{thebibliography}{53}
\bibitem [{SM()}]{SM}%
  \BibitemOpen
  \href@noop {} {\bibinfo {title} {See {S}upplemental {M}aterial at [url will
  be inserted by publisher] for further experimental data, detailed derivation
  of the extensions, and entropy estimation with variable number of
  particles.}}\BibitemShut {Stop}%
\end{thebibliography}
\end{document}